\documentclass[dvipsnames, twocolumn]{aastex631}

\usepackage{amsmath}
\usepackage{graphicx}

\newcommand{\ee}[1]{\mbox{${} \times 10^{#1}$}}
\newcommand{\eten}[1]{\mbox{$10^{#1}$}}
\newcommand{\lsun}{\mbox{L$_\odot$}}
\newcommand{\msun}{\mbox{M$_\odot$}}

\newcommand{\degree}{\mbox{$^{\circ}$}}

\newcommand{\um}{$\mathrm{\mu}$m}
\newcommand{\cmcm}{cm$^{-2}$}

\newcommand{\water}{H$_2$O}
\newcommand{\co}{$^{12}$CO}
\newcommand{\coo}{$\mathrm{^{13}CO}$}
\newcommand{\cooo}{C$^{18}$O}

\newcommand{\cotwo}{CO$_2$}
\newcommand{\thcotwo}{$^{13}$CO$_2$}
\newcommand{\hh}{\mbox{{\rm H}$_2$}}

\newcommand{\cchh}{\mbox{{\rm C}$_2$\rm{H}$_2$}}

\def\jwst{\emph{JWST}}
\def\spitzer{\emph{Spitzer}}

\begin{document}

\title{MINDS. JWST-MIRI Reveals a Dynamic Gas-Rich Inner Disk Inside the Cavity of SY Cha}

\correspondingauthor{Kamber R. Schwarz}
\email{schwarz@mpia.de}

\author[0000-0002-0786-7307]{Kamber R. Schwarz}
\affil{Max-Planck-Institut f\"{u}r Astronomie (MPIA), K\"{o}nigstuhl 17, 69117 Heidelberg, Germany}

\author[0000-0002-1493-300X]{Thomas Henning}
\affil{Max-Planck-Institut f\"{u}r Astronomie (MPIA), K\"{o}nigstuhl 17, 69117 Heidelberg, Germany}

\author[0000-0002-0101-8814]{Valentin Christiaens}
\affil{STAR Institute, Universit\'e de Li\`ege, All\'ee du Six Ao\^ut 19c, 4000 Li\`ege, Belgium}

\author[0000-0002-1257-7742]{Danny Gasman}
\affil{Institute of Astronomy, KU Leuven, Celestijnenlaan 200D, 3001 Leuven, Belgium}

\author[0000-0001-9992-4067]{Matthias Samland}
\affil{Max-Planck-Institut f\"{u}r Astronomie (MPIA), K\"{o}nigstuhl 17, 69117 Heidelberg, Germany}

\author[0000-0002-8545-6175]{Giulia Perotti}
\affil{Max-Planck-Institut f\"{u}r Astronomie (MPIA), K\"{o}nigstuhl 17, 69117 Heidelberg, Germany}

\author{Hyerin Jang}
\affil{Department of Astrophysics/IMAPP, Radboud University, PO Box 9010, 6500 GL Nijmegen, The Netherlands}

\author[0000-0002-4022-4899]{Sierra L. Grant}
\affil{Max-Planck Institut f\"{u}r Extraterrestrische Physik (MPE), Giessenbachstr. 1, 85748, Garching, Germany}

\author{Beno\^{i}t Tabone}
\affil{Universit\'e Paris-Saclay, CNRS, Institut d’Astrophysique Spatiale, 91405, Orsay, France}

\author[0000-0001-9526-9499]{Maria Morales-Calder\'on}
\affil{Centro de Astrobiolog\'ia (CAB), CSIC-INTA, ESAC Campus, Camino Bajo del Castillo s/n, 28692 Villanueva de la Ca\~nada,
Madrid, Spain}

\author[0000-0001-7455-5349]{Inga Kamp}
\affil{Kapteyn Astronomical Institute, Rijksuniversiteit Groningen, Postbus 800, 9700AV Groningen, The Netherlands}

\author[0000-0001-7591-1907]{Ewine F. van Dishoeck}
\affil{Leiden Observatory, Leiden University, 2300 RA Leiden, the Netherlands}
\affil{Max-Planck Institut f\"{u}r Extraterrestrische Physik (MPE), Giessenbachstr. 1, 85748, Garching, Germany}


\author[0000-0001-9818-0588]{Manuel G\"udel}
\affil{Dept. of Astrophysics, University of Vienna, T\"urkenschanzstr. 17, A-1180 Vienna, Austria}
\affil{Max-Planck-Institut f\"{u}r Astronomie (MPIA), K\"{o}nigstuhl 17, 69117 Heidelberg, Germany}
\affil{ETH Z\"urich, Institute for Particle Physics and Astrophysics, Wolfgang-Pauli-Str. 27, 8093 Z\"urich, Switzerland}

\author{Pierre-Olivier Lagage}
\affil{Universit\'e Paris-Saclay, Universit\'e Paris Cit\'e, CEA, CNRS, AIM, F-91191 Gif-sur-Yvette, France}


\author[0000-0002-5971-9242]{David Barrado}
\affil{Centro de Astrobiolog\'ia (CAB), CSIC-INTA, ESAC Campus, Camino Bajo del Castillo s/n, 28692 Villanueva de la Ca\~nada, Madrid, Spain}

\author[0000-0001-8876-6614]{Alessio Caratti o Garatti}
\affil{INAF – Osservatorio Astronomico di Capodimonte, Salita Moiariello 16, 80131 Napoli, Italy}
\affil{Dublin Institute for Advanced Studies, 31 Fitzwilliam Place, D02 XF86 Dublin, Ireland}

\author[0000-0001-9250-1547]{Adrian M. Glauser}
\affil{ETH Z\"urich, Institute for Particle Physics and Astrophysics, Wolfgang-Pauli-Str. 27, 8093 Z\"urich, Switzerland}

\author[0000-0002-2110-1068]{Tom P. Ray}
\affil{Dublin Institute for Advanced Studies, 31 Fitzwilliam Place, D02 XF86 Dublin, Ireland}

\author[0000-0002-1368-3109]{Bart Vandenbussche}
\affil{Institute of Astronomy, KU Leuven, Celestijnenlaan 200D, 3001 Leuven, Belgium}

\author[0000-0002-5462-9387]{L. B. F. M. Waters}
\affil{Department of Astrophysics/IMAPP, Radboud University, PO Box 9010, 6500 GL Nijmegen, The Netherlands}
\affil{SRON Netherlands Institute for Space Research, Niels Bohrweg 4, NL-2333 CA Leiden, the Netherlands}


\author[0000-0001-8407-4020]{Aditya M. Arabhavi}
\affil{Kapteyn Astronomical Institute, Rijksuniversiteit Groningen, Postbus 800, 9700AV Groningen, The Netherlands}

\author[0000-0003-0386-2178]{Jayatee Kanwar}
\affil{Kapteyn Astronomical Institute, Rijksuniversiteit Groningen, Postbus 800, 9700AV Groningen, The Netherlands}
\affil{Space Research Institute, Austrian Academy of Sciences, Schmiedlstr. 6, A-8042, Graz, Austria}
\affil{TU Graz, Fakultät für Mathematik, Physik und Geodäsie, Petersgasse 16 8010 Graz, Austria}

\author[0000-0003-3747-7120]{G\"oran Olofsson}
\affil{Department of Astronomy, Stockholm University, AlbaNova University Center, 10691 Stockholm, Sweden}

\author[0000-0002-0100-1297]{Donna Rodgers-Lee}
\affil{Dublin Institute for Advanced Studies, 31 Fitzwilliam Place, D02 XF86 Dublin, Ireland}

\author{J\"urgen Schreiber}
\affil{Max-Planck-Institut f\"{u}r Astronomie (MPIA), K\"{o}nigstuhl 17, 69117 Heidelberg, Germany}

\author[0000-0002-7935-7445]{Milou Temmink}
\affil{Leiden Observatory, Leiden University, 2300 RA Leiden, the Netherlands}


\begin{abstract}
SY Cha is a T Tauri star surrounded by a protoplanetary disk with a large cavity seen in the millimeter continuum but has the spectral energy distribution (SED) of a full disk. Here we report the first results from \jwst-MIRI Medium Resolution Spectrometer (MRS) observations taken as part of the MIRI mid-INfrared Disk Survey (MINDS) GTO Program. The much improved resolution and sensitivity of MIRI-MRS compared to \spitzer\ enables a robust analysis of the previously detected \water, CO, HCN, and \cotwo\ emission as well as a marginal detection of \cchh. We also report the first robust detection of mid-infrared OH and ro-vibrational CO emission in this source. The derived molecular column densities reveal the inner disk of SY Cha to be rich in both oxygen and carbon bearing molecules. This is in contrast to PDS 70, another protoplanetary disk with a large cavity observed with \jwst, which displays much weaker line emission. In the SY Cha disk, the continuum, and potentially the line, flux varies substantially between the new \jwst\ observations and archival \spitzer\ observations, indicative of a highly dynamic inner disk.
\end{abstract}

\section{Introduction} \label{sec:intro}
SY Cha is a $1-2$ million year old protoplanetary disk in the Chamaeleon I star-forming region at a distance of 180.7 pc \citep{Gaia21, Galli21}. Recent ALMA observations revealed a large cavity with a radius 70 au in the millimeter continuum emission\citep{Orihara23}. 
Such cavities are a hallmark of transtion disks.
Transition disks were first identified as a sub-class of protoplanetary disks based on the comparatively small excess of emission relative to a stellar blackbody in the near infrared \citep{Strom89}. This was interpreted as due to a deficiency of dust in the inner regions of the disk, which was later confirmed by imaging in the sub-millimeter \citep{Calvet02,Hughes07}.
During the \spitzer\ era, the term pre-transitional disks was coined to describe disks with a deficit of emission from 5-20 \um\ while still displaying excess emission in the 2-5 \um\ range, interpreted as arising from an optically thick inner disk \citep{Brown07,Espaillat07}.

Many transition disks have now been identified, both from SEDs and imaging \citep{Andrews11,Espaillat14,Ercolano17,Francis20}. 
While the term `transition disk' was first coined based on SEDs, here we use the more general definition of any protoplanetary disk with a large dust cavity, regardless of whether a system would be classified as transitional or pre-transitional based on its SED alone \citep[e.g.,][]{Espaillat14,vanderMarel23}.
The mismatch between SED and millimeter imaging can be due to, e.g., a misaligned inner disk or a millimeter grain poor cavity which still contains micron-sized grains \citep{vanderMarel18,Villenave19}.
The mechanisms through which the inner disk is cleared of millimeter grains remain unclear. Potential explanations include clearing by photoevaporative winds, inward drift of large grains, and gap clearing by a low-mass companion \citep{Lin79,Birnstiel12,Alexander14,Bae23,Pascucci23}.

The median accretion rate for stars with transition disks is comparable to that of full protoplanetary disks, i.e., those without large inner dust cavities, indicating gas is still present at small radii \citep{SiciliaAguilar10}. 
Photoevaporative winds are disfavored as the cavity forming mechanism for disks with large accretion rates \citep{vanderMarel18}. 
Many transition disks have gas cavities which are smaller than the associated dust cavity, which is expected in the planet-disk scenario \citep[e.g.,][]{Carmona14,Carmona17}.
Two such disks, PDS 70 and HD 169142, have been conclusively shown to host planets in their dust cavites \citep{Muller18,Keppler18,Haffert19,Gratton19,Hammond23}. Tentative evidence for planets within the inner cavity of transition disks exists for several additional systems \citep[e.g.,][]{Long22,Currie22,Stadler23}. Whether all transition disks are formed by planet clearing or multiple mechanisms are at play remains unclear.

Observations of transition disks with the \spitzer\ InfraRed Spectrograph (IRS) reveal that mid-IR variability on timescales of several years is common, with two general types of variability \citep{Espaillat11}. The first is when emission at short and long wavelengths varies inversely in a ``seesaw'' behavior. Such variability could be caused by unseen planets warping the inner disk \citep{Muzerolle09}. 
A more uniform intensity shift at all mid-IR wavelengths has also been observed in transition disks, and is likely due to a change in incident flux from the central star \citep{Espaillat11}.

Spectral indices, a measure of the ratio of the flux density for two wavelength ranges, denoted $n$, are used to characterize the SEDs of protoplanetary disks \citep[e.g.,][]{Kessler06,Furlan06}.
\citet{Pontoppidan10} found that the five systems classified as transition disks in their \spitzer\ sample, based on their infrared spectral index, $n_{13-30}$, did not exhibit molecular line emission other than \hh\ down to the 1\% line-to-continuum level. 
Subsequent work by \citet{Banzatti20} found an inverse correlation between the outer radius of the millimeter dust disk as observed by ALMA and the luminosity of \water\ lines in the mid-IR. They further speculated that disks with large inner cavities may have weaker \water\ lines compared to other large disks, since icy pebbles are prevented from drifting inward. 
However, they note additional data is needed to confirm this hypothesis. It has also been suggested that even when pebbles are stopped from moving inward, micron-size particles well coupled to the gas can transport water inward \citep{Potapov18,Perotti23}.

Here we report our findings on the inner disk of SY Cha, based on JWST observations. 
SY Cha was observed as part of the as part of the MIRI mid-INfrared Disk Survey (MINDS) GTO Program (PID: 1282, PI: T. Henning). Previous results from the MINDS program include the first detection of \thcotwo\ in a protoplanetary disk \citep{Grant23}, rich hydrocarbon chemistry in a disk around a very low-mass star \citep{Tabone23}, abundant water but a sub-stellar C/O ratio in a full protoplanetary disk \citep{Gasman23b}, and water in the inner disk of a transition disk known to host planets in its cavity \citep{Perotti23}. An overview of the full MINDS sample is given by \citet{Kamp23} and Henning et al.\,(in prep).

SY Cha would not be classified as a transition disk based on its infrared spectral index of $n_{13-30} = -0.17$ \citep{Banzatti20}. However, recent ALMA observations reveal a ring of continuum emission centered at 
$0\farcs{559}$ (101 au) with a FWHM of $0\farcs{22}$ (40 au), an inner gap, and unresolved emission centered on the star with a radius no larger than $0\farcs{025}$ \citep[2.3 au,][]{Orihara23}. 
This small scale component could be due to an inner disk, free-free emission, or an ionized wind \citep{Francis20}. 
Other source properties are listed in Table~\ref{sourcetab}. The \coo\ and \cooo\ $J=2-1$ emission also appears to be ring like, though with an inner radius smaller than that of the dust ring. Additionally, near the central star the \co\ emission displays a deviation from Keplerian rotation, which \citet{Orihara23} posit could be due to either a warped inner disk or a radial flow such as a wind. 
Interestingly, the CO ro-vibrational lines at 4.7 \um, which often include a disk wind component, were not detected toward SY Cha using the high spectral resolution CRIRES instrument on the VLT \citep{Brown13}.
Either a highly inclined inner disk or the presence of gas (and thus also small dust grains coupled to the gas) inside the millimeter cavity would explain why SY Cha's SED does not include the classic signature of a transition disk. \citet{Orihara23} found models with a highly inclined inner disk to be consistent with their ALMA observations, though more modeling work is needed to provide a definitive result. 
\spitzer\ detected mid-infrared emission originating from the inner disk for \water, HCN, \cchh, and \cotwo, while only upper limits for OH were reported \citep{Salyk11,Banzatti20}. The much improved sensitivity and spectral resolution of MIRI-MRS compared to \spitzer-IRS enables a more detailed analysis of the mid-IR line emission, particularly in the 5-10 \um\ range, which was not covered by \spitzer-IRS's high resolution mode.

\begin{deluxetable}{ll}
\tablewidth{\columnwidth}
\tablecolumns{2}
\tablecaption{Properties of SY Cha}
\label{sourcetab}
\tablehead{
\colhead{Parameter} & \colhead{Value}  
}
\startdata
Stellar mass & 0.70 \msun\ \\
Stellar luminosity & 0.55 \lsun\ \\
Effective temperature & 4030 K \\
Accretion luminosity & 7.39\ee{-3} \lsun\ yr$^{-1}$ \\
Mass accretion rate & 6.6\ee{-10} \msun\ yr$^{-1}$ \\
Spectral type & K5Ve \\
Distance & 180.7 pc \\
Disk inclination & 51.1\degree\ \\
\enddata
\tablecomments{
Spectral type classification from \citet{Frasca15}. Disk inclination from \citet{Orihara23}. All other values are from \citet{Manara23}, using updated data from \citet{Gaia21} and stellar models from \citet{Feiden16}.
}
\end{deluxetable}

\section{Data Reduction \& Analysis}\label{reduction}
\subsection{Observations and Data Reduction}
SY Cha was observed with the Mid-InfraRed Instrument (MIRI)\footnote{\citet{Rieke15,Wright15}} in Medium Resolution Spectroscopy (MRS)\footnote{\citet{Wells15}} mode on 8 August, 2022 as part of the MINDS GTO Program (PID:
1282, PI: T. Henning). The total exposure time was 3696 seconds using a four-point 
dither pattern in the positive direction. 

\begin{figure*}
    \centering
    \gridline{\fig{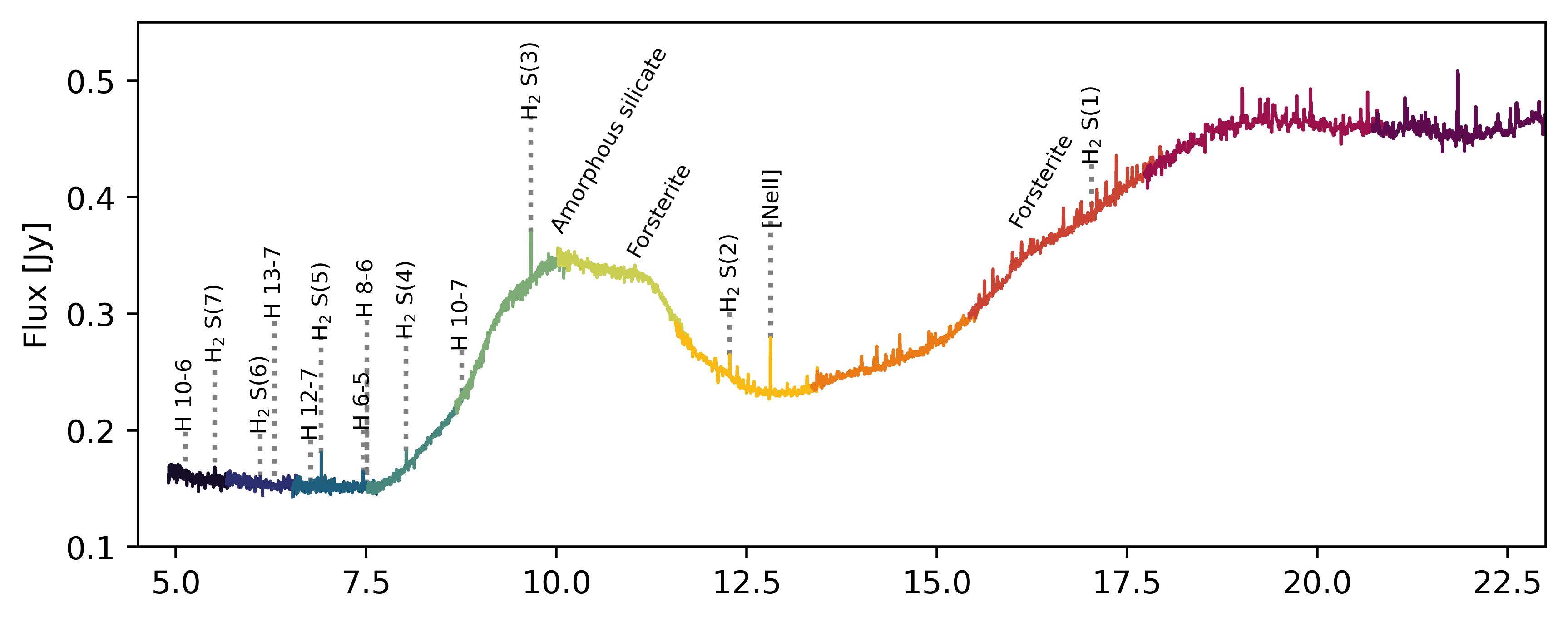}{0.9\textwidth}{}}
    \vspace{-0.8cm}
    \gridline{\fig{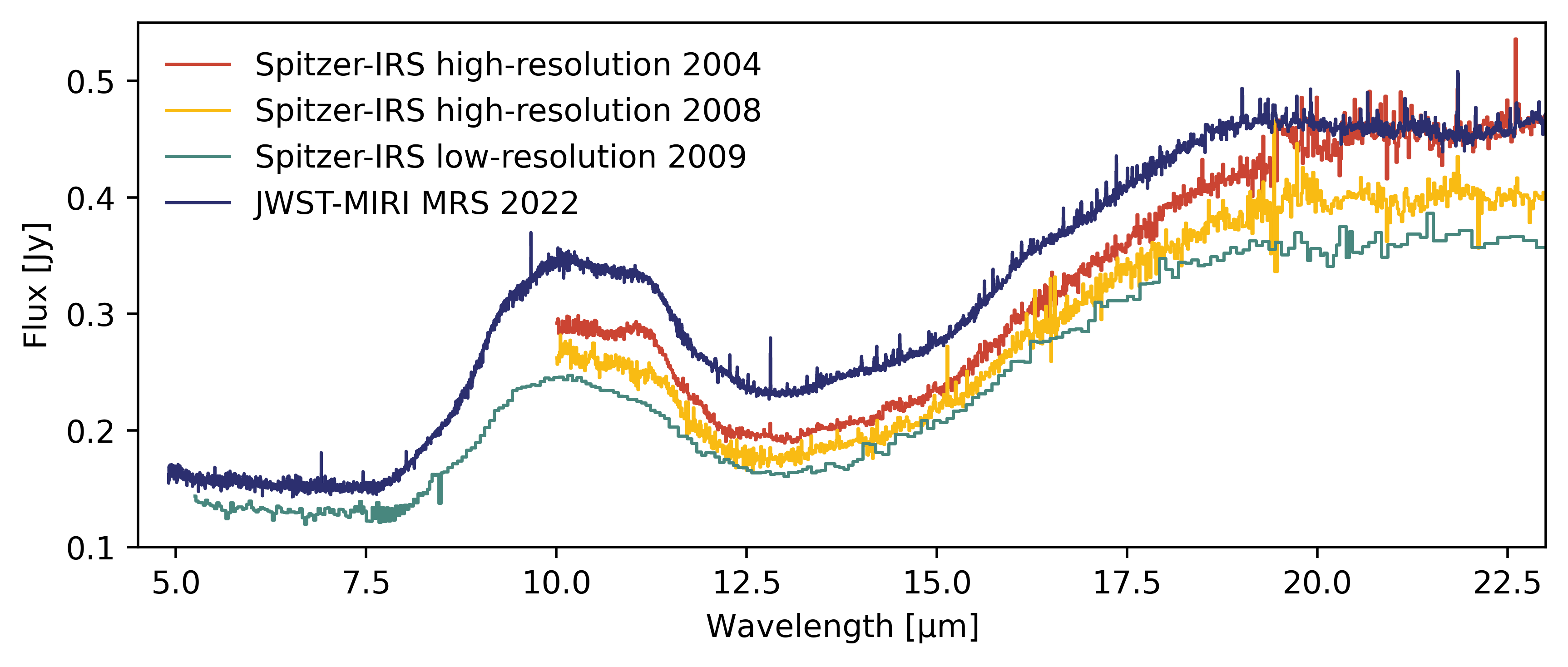}{0.9\textwidth}{}}
    \vspace{-0.8cm}
    \caption{Top: \jwst-MIRI MRS spectrum of SY Cha. Coloring designates the short, medium, and long sub-bands in the four integral field units. Major atomic and \hh\ lines, as well as dust features, are labeled. Bottom: Mid-IR spectrum of SY Cha as observed by \spitzer-IRS in high resolution and low resolution mode compared to \jwst-MIRI MRS. \spitzer\ observations taken from CASSIS \citep{CASSIS1,CASSIS2}. }
    \label{fullspec}
\end{figure*}

The data were processed using version 1.9.4 of the \jwst\ pipeline \citep{Bushouse23}.
Bespoke routines based on the VIP package were used for background subtraction, bad pixel correction after stage 2, and spectrum extraction \citep{VIP1,VIP2}. At the Spec2 stage the standard flat fielding and fringe correction from the \jwst\ pipeline were used. 
The background was then created using the minimum values among the four dither positions and smoothed with a median filter. 
Spectra were extracted from the final cubes using apertures twice the full width half max in size centered on the source centroid, identified with a 2D Gaussian fit. 
After extraction, residual fringe correction was carried out for each wavelength band. The final spectrum is presented in Figure~\ref{fullspec}.

Visual inspection of the data cubes reveals the \hh\ and [NeII] emission to be extended, while the rest of the line emission and the continuum is spatially unresolved. The above data reduction was not optimized for extended emission. Analysis of the extended \hh\ emission, likely indicating the molecular component of a wind, will be presented in a separate work and is not discussed further here.

\subsection{Continuum Emission}\label{results:cont}
The overall spectrum shows stronger emission than observed by \spitzer\ in either low-resolution or high-resolution mode, with the \spitzer\ high resolution observations (2004-08-31 and 2008-05-01) sitting between the low resolution (2009-04-09) and MIRI observations (2022-08-08, see Figure~\ref{fullspec}). 
The largest differences are seen between the new MIRI observations and the low resolution \spitzer\ observations from 2009, with the MIRI spectrum roughly 10\% stronger at 6 \um, 40\% at 10 \um, and 30\% at 20 \um.
The spectro-photometric accuracy is 2-10\% for the \spitzer\ IRS and $5.6\pm0.7\%$ for the \jwst-MIRI MRS \citep{Furlan06,Watson09,Argyriou23}. The increased spatial resolution of \jwst\ lessens the chance of contaminating emission from other objects. Such contamination in the \spitzer\ data would result in a higher flux relative to \jwst, the opposite of what we observe.
While the difference in flux at short wavelengths is within the calibration uncertainty, the difference beyond 10 \um\ is not.

Transition disks are known to be variable in the IR. The level of variability seen between the \spitzer\ and \jwst\ observations of SY Cha is similar to the variability of transition and pre-transition disks observed over multiple epochs with \spitzer\ by \citet{Espaillat11}. That the variability is greater at longer wavelengths points to changes in the inner disk geometry as the source of the variability in SY Cha. A more nuanced understanding of the cause of the mid-IR variability in SY Cha requires detailed radiative transfer modeling, which will be presented in subsequent work. 

\subsection{Fitting Atomic Emission}\label{results:atom}
\begin{figure}[!h]
    \centering
    \includegraphics[width=\columnwidth]{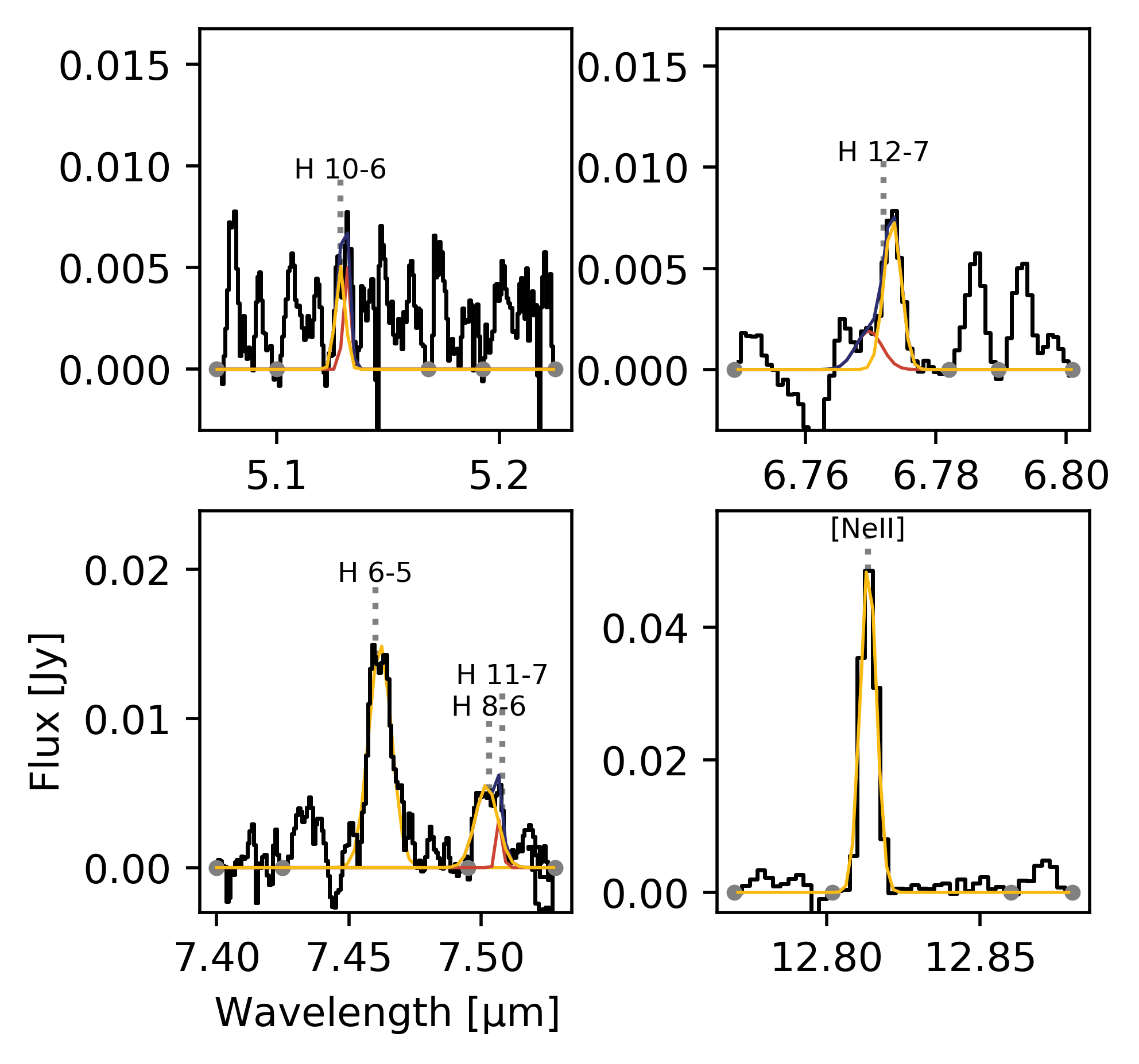}
    \caption{Zoom in of continuum subtracted spectra at the brightest atomic lines. Grey dots denote wavelengths used for continuum subtraction. Yellow line shows the best-fit Gaussian for the atomic line. When applicable, red and dark blue lines show the additional Gaussian fit to overlapping lines and the combined fit respectively.}
    \label{atoms}
\end{figure}

The strongest atomic line features are shown in Figure~\ref{atoms}. To determine the total flux from each transition, we fit a Gaussian line profile using the Markov Chain Monte Carlo package \texttt{emcee} \citep{emcee}. Because many of the atomic lines overlap with nearby transitions, a two component Gaussian is used. Only the Gaussian fit to the atomic transition is used when integrating to get the total flux. Each Gaussian fit is integrated over the respective wavelength range shown in Figure~\ref{atoms}.
For subtraction purposes, the continuum is defined by selecting apparently line-free pixels (See App.~\ref{app:contsub}) and then interpolating with a spline fit. 
The spline fit to these points is then subtracted from the spectrum. 
The rms noise, $\sigma$, of the continuum subtracted spectrum is calculated by taking the standard deviation in a line-free wavelength range. Values for $\sigma$ are given in Table~\ref{atomicprops}. 
Best-fit models are identified using a modified reduced $\chi^2$ fitting:
\begin{equation}
\chi^2 = \frac{1}{N} \sum_{i=1}^N \frac{(F_{obs,i}-F_{mod,i})^2}{\sigma^2}
\end{equation}
where $N$ is the number of data points used in the fitting, $F_{obs}$ and $F_{mod}$ are the observed and model line flux, respectively, and $\sigma$ is the root mean squared noise calculated in a nearby line-free region. 

\begin{deluxetable*}{ccccccc}
\tablewidth{0pt}
\tablecolumns{2}
\tablecaption{Integrated flux for Gaussian fits to atomic lines}
\label{atomicprops}
\tablehead{
\colhead{Transition} & \colhead{Line Center} & \colhead{$\mathrm{\sigma}$}  & \colhead{Integrated Flux} & \colhead{Continuum} & \colhead{rms} & \colhead{$\mathrm{\chi^2}$}  \\  
 & \colhead{(\um)} & \colhead{10$^{-3}$(\um)} & \colhead{10$^{-15}$(erg s$^{-1}$ cm$^{-2}$)} & \colhead{(\um)} & \colhead{$10^{-27}$(erg s$^{-1}$ cm$^{-2}$ Hz$^{-1}$)} & \colhead{}
}
\startdata 
HI 10-6 & 5.129  & 1.2 & 1.98 & 5.096-5.102  & 1.52 & 25.5\\
HI 12-7 & 6.772 & 1.3 & 1.62 & 6.777-6.782 & 0.46  & 26.9\\
HI 6-5 & 7.460 & 4.3 & 8.17 & 7.49-7.4954 & 2.65  & 20.3\\
HI 8-6 & 7.503 & 4.2 & 1.58 & 7.49-7.4954 & 2.65  & 1.24\\
HI 11-7& 7.508  & 1.1 & 0.48 & 7.49-7.4954 & 2.65  & 1.24\\
\enddata 
\end{deluxetable*}

\subsection{Fitting Molecular Emission}
\begin{figure*}
    \centering
    \includegraphics[]{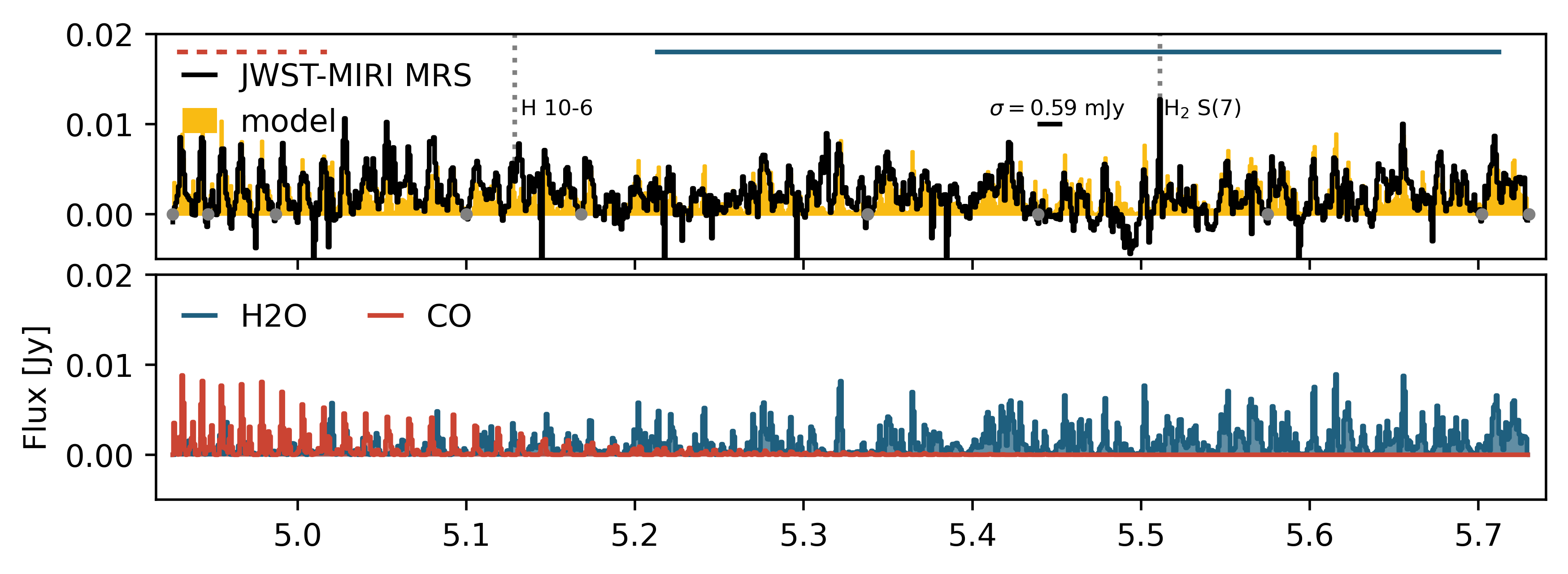}
    \includegraphics[]{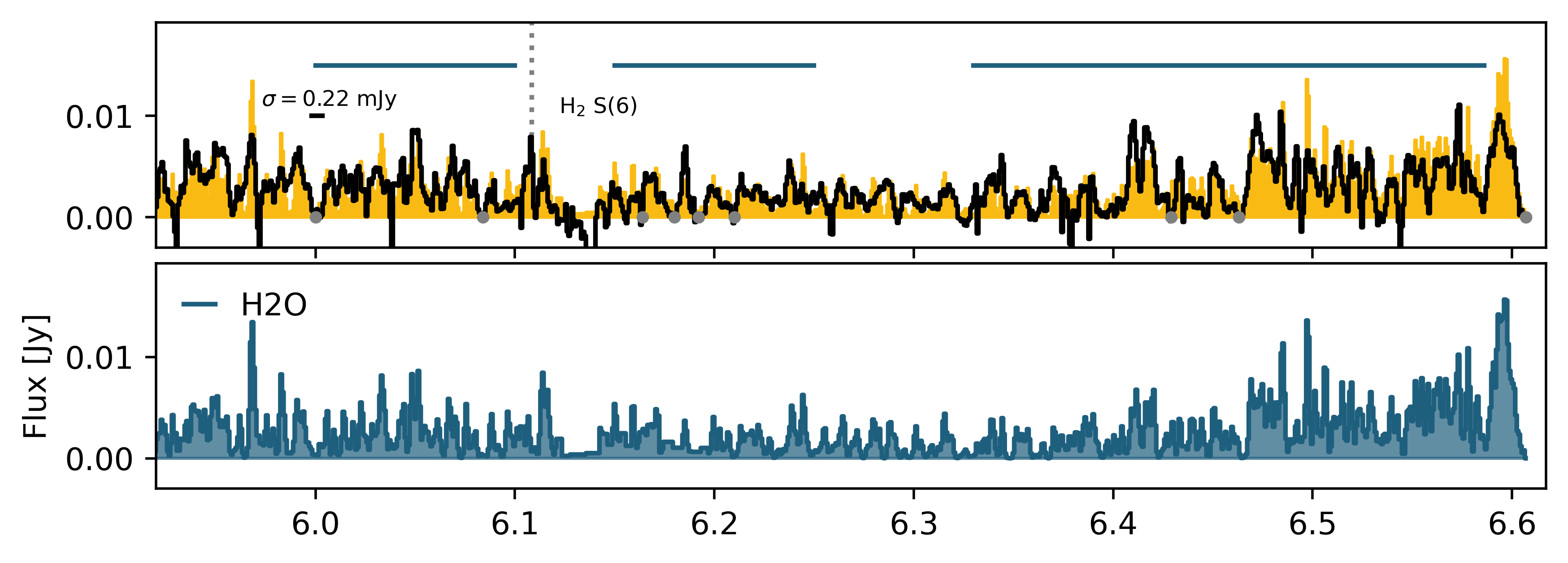}
    \includegraphics[]{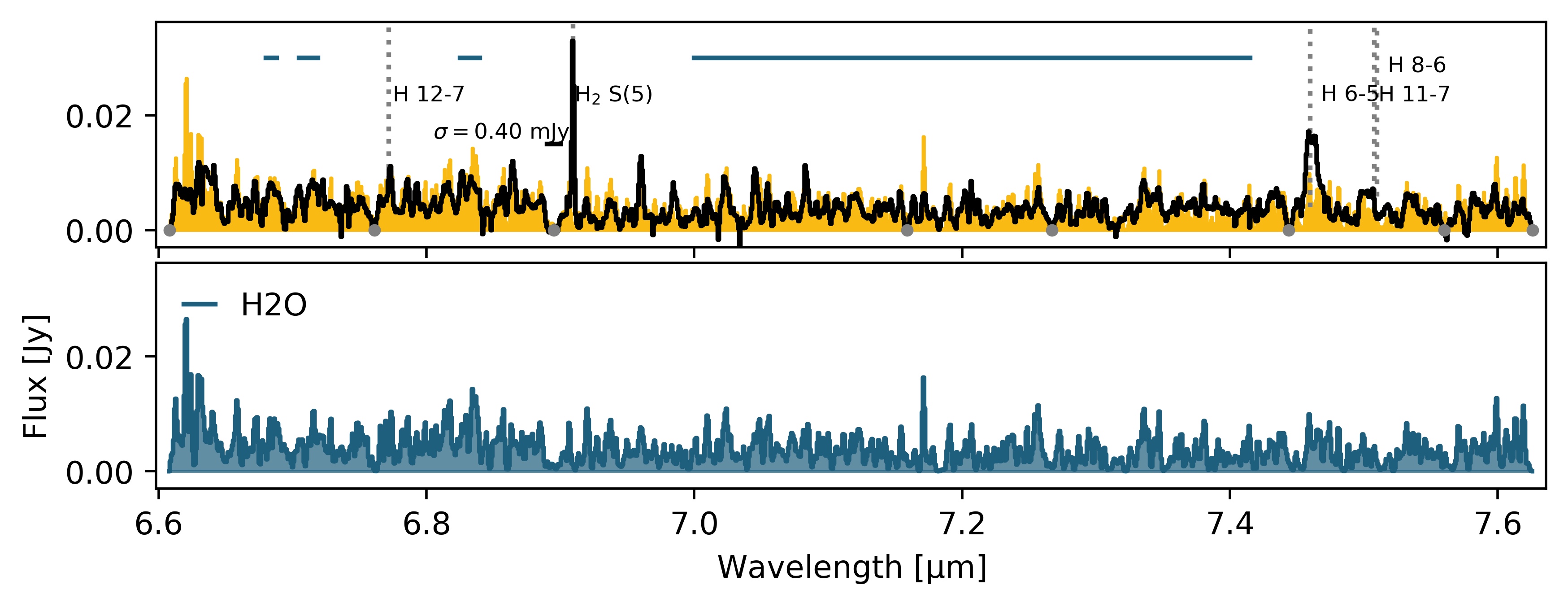}
    \caption{
Top: Continuum subtracted observations (black) and cumulative model adding the contributions from each molecule (yellow) from 4.9 to 7.6 \um. Grey dots denote wavelengths used for continuum subtraction. Black horizontal lines indicate the line-free region used to calculate the rms noise. Other horizontal lines indicate the wavelength range considered when fitting each molecule. Bottom: Individual LTE slab models for each molecule.}
    \label{slabfits1}
\end{figure*}

\begin{figure*}
    \centering
    \includegraphics[]{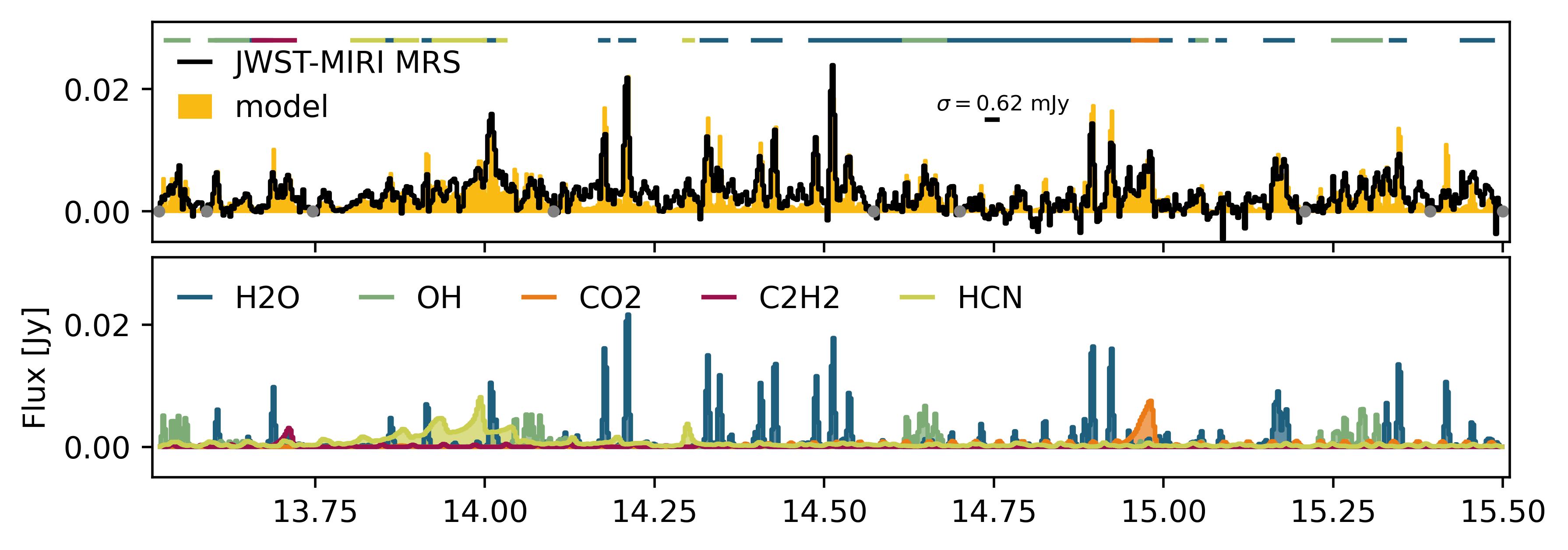}
    \includegraphics[]{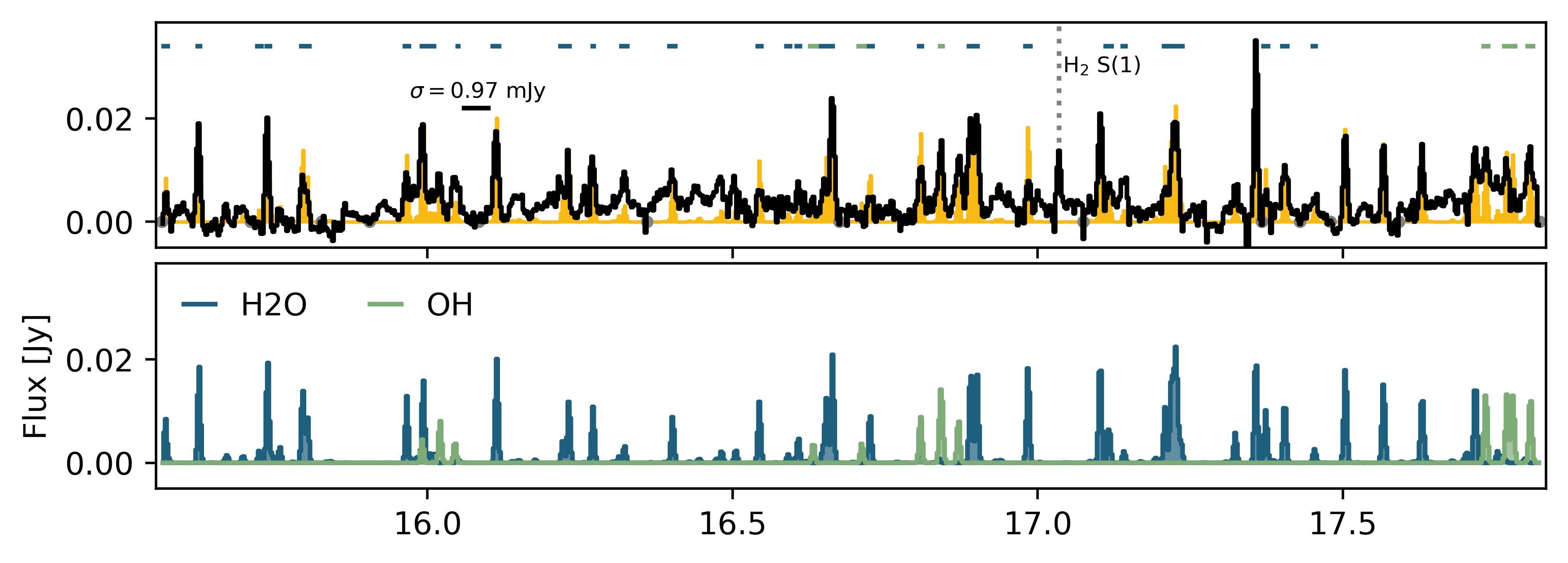}
    \includegraphics[]{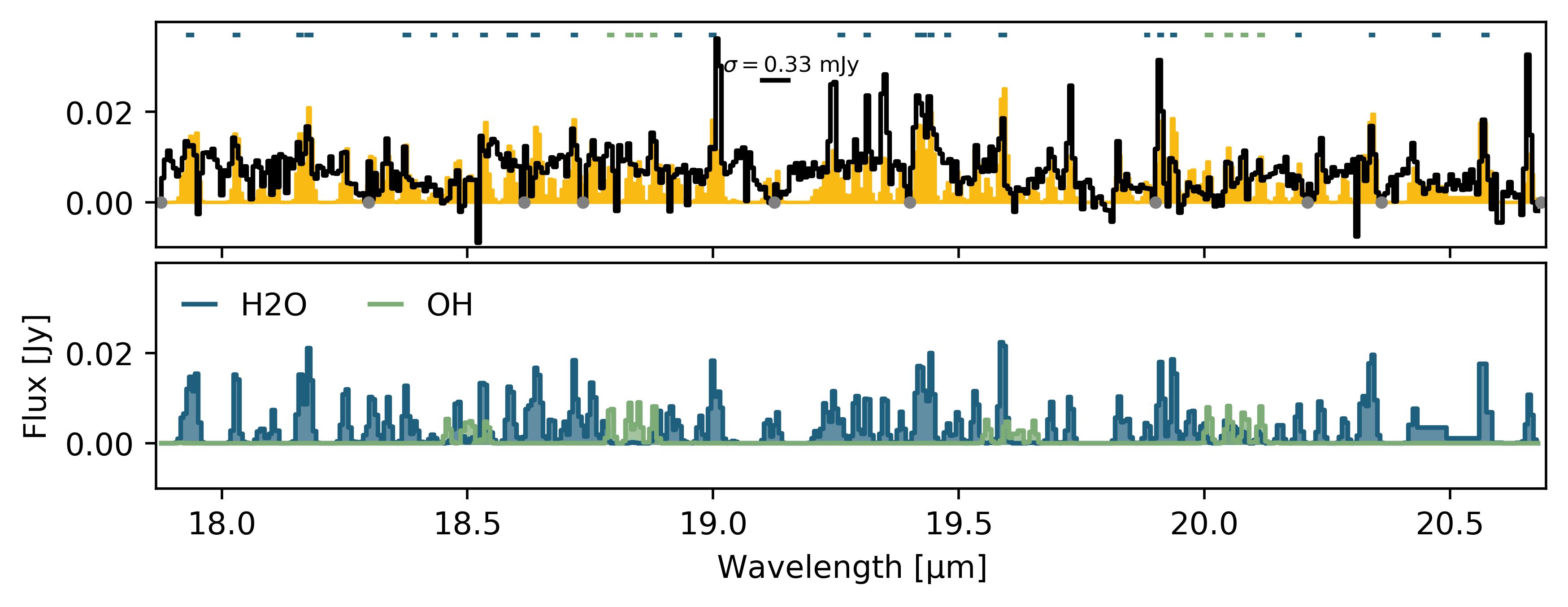}
    \caption{Top: Continuum subtracted observations (black) and cumulative model adding the contributions from each molecule (yellow) from 13.5 to 20.7 \um. Grey dots denote wavelengths used for continuum subtraction. Black horizontal lines indicate the line-free region used to calculate the rms noise. Other horizontal lines indicate the wavelength range considered when fitting each molecule. Bottom: Individual LTE slab models for each molecule.}
    \label{slabfits2}
\end{figure*}

We characterize the observed molecular emission by fitting local thermodynamic equilibrium (LTE) 0D slab models, as presented by \citet{Tabone23}, to the individual sub-bands. Each model is a slab of gas with a uniform temperature and density and no implicit spatial information. The LTE assumption is valid when all energy levels of a given species are well characterized by a single excitation temperature. 
Continuum subtraction and noise estimation is carried out in the same way as for the atomic emission (Section~\ref{results:atom}). The points used for continuum subtraction are listed in App.~\ref{app:contsub}.  The wavelength range used to calculate the noise and the resulting values are shown in Figures~\ref{slabfits1} \& \ref{slabfits2}.

When generating the slab models, the intrinsic line profile is assumed to be a Gaussian.
We include mutual shielding from closely spaced transitions of the same molecule following the method described by \citet{Tabone23}:
\begin{equation}
\tau(\lambda)  = \sum_{i} \tau_{0,i}  e^{-(\lambda-\lambda_{0,i})^2/2 \sigma_{\lambda}^2},
\end{equation}
where  $\lambda_{0,i}$ is the rest wavelength of a given line $i$, $\sigma_{\lambda}$ is the intrinsic broadening of the line in microns, assumed to be equivalent to $\sigma_{\lambda} = 2$ km s$^{-1}$ \citep{Salyk11b} and $\Delta V = 4.7$ km s$^{-1}$, and $\tau_{0,i}$ is the optical depth at line center:
\begin{equation}
\tau_{0}  = \sqrt{\frac{\ln 2}{\pi}} \frac{A_{ul} N \lambda_0^3}{ 4 \pi  \Delta V} (x_{l} \frac{g_u}{g_l} - x_{u}).
\end{equation}
Here, $x_{u}$ and $x_{l}$ are the level population of the upper and lower states respectively, $g_u$ and $g_l$ are their respective
statistical weights, $A_{ul}$ is the Einstein A coefficient, giving the rate of spontaneous emission, and $N$ is the total column density of the molecule.

To find the best-fit for each molecule a grid of models is generated with three free parameters: column density $N$ (grid spacing $\Delta \log N = 0.2$ dex), excitation temperature $T$ ($\Delta T = 50$ K), and emitting area $\pi R^2$ ($\Delta \log R = 0.003$ dex):
\begin{equation}
F(\lambda) = \pi \left( \frac{R}{d} \right)^2 B_{\nu}(T) (1-e^{-\tau(\lambda)}),
\end{equation}
where d is the assumed distance to SY Cha of 180.7 pc.
Note that $R$ does not correspond to a physical radius in the disk, but rather is the radius of a circle with the same area as the derived emitting area. As such, it $R$ can be considered a lower limit on the physical radius of the emission, though one that is not motivated by any physical model.
The resulting model spectra are convolved to the resolution of the corresponding MRS band, using the average values reported by \citet{Jones23} for each band, then resampled to the same wavelength grid as the observations using the \textit{SpectRes} package \citep{Carnall17}.

For a given sub-band, the spectral lines are fit for one molecule at a time, following the method of \citet{Grant23}. The goodness of fit for a given model is determined by comparing to the observations at each wavelength using a $\chi^2$ routine. The $\chi^2$ fitting is restricted to regions where emission from the given molecule is expected and there is no expected strong emission from other species based on initial slab models using typical emitting temperatures for each species. 
Additionally, pixels where the continuum-subtracted flux is less than -5 mJy are excluded. This excludes both potential absorption features and single pixels with anomalously low values.

After a best-fit model for a given molecule has been identified, this model is subtracted from the spectrum before fitting the next species.
Molecules are fit in the following order: 
\water, OH, CO,
\cotwo, HCN,
\cchh. 
After obtaining an initial fit for all species, we rerun the fit for \water\ with the best-fit models from all other species subtracted in order to account for any small contributions from other species overlapping with the water emission. 
This second round of fitting results in only small variations, within the reported uncertainty, compared to the first round, demonstrating that the emission from other species has only a minor impact on our fitting routine, at least for species with strong lines. For weaker lines, it is possible that this method leads to an underestimation of the line flux.
Molecular fits are only run when the molecule is expected to emit within the sub-band. The continuum subtracted spectra and best-fit models for the fitted regions are shown in Figures~\ref{slabfits1} \& \ref{slabfits2} and the $\chi^2$ maps for the model grid are shown in Figures~\ref{chi2water}-\ref{chi2other}. 

\section{Results}
The overall MIRI-MRS spectrum of SY Cha is characterized by a strong 10 \um\ amorphous silicate feature in emission as well as evidence of forsterite.
Following the method of \citet{vanBoekel05} we measure the ratio of the flux at 11.3 and 9.8 \um, which is correlated with the strength of the silicate feature. We find a low 11.3/9.8 \um\ ratio of $\sim 0.7$. This is indicative of the presence of warm silicate grains which have not undergone a large amount of processing. This emission originates in the inner disk and demonstrates the presence of a dusty disk at small radii.

\subsection{Atomic Hydrogen Emission}
Atomic hydrogen emission at mid-IR wavelengths arises primarily from the hot disk atmosphere, though jets and outflows can also contribute. We detect four hydrogen recombination lines shortward of 10 \um, with the strongest being the HI (6-5) line at $\sim$7.5 \um. We do not detect the HI (7-6) line, which is often used as an accretion tracer, \citep[e.g.,][]{Rigliaco15,Beuther23}, though it may be masked by the presence of multiple strong \water\ lines in the vicinity. 

\subsection{Molecular Emission}
Figures~\ref{slabfits1} \& \ref{slabfits2} show the continuum subtracted spectra as well as the best-fit slab models. The best-fit model parameters are given in Table~\ref{slabprops}. 
After continuum subtraction, the MIRI-MRS spectrum of SY Cha is dominated by ro-vibrational and pure rotational \water\ vapor emission, which is seen in all sub-bands (Figures~\ref{slabfits1} \& \ref{slabfits2}). As with observations of other disks observed with MIRI-MRS, e.g., \citet{Kospal23,Banzatti23,Gasman23b}, the lines at longer wavelengths preferentially trace cooler gas (Table~\ref{slabprops}). In addition to \water, we detect CO, \cchh, HCN, \cotwo, and OH.
The best-fit column densities for \water\ are high compared to most other disks observed with MIRI, where typical column densities are of order $10^{18}$ cm$^{-2}$, while the column densities for the carbon bearing species are on the low end \citep{Grant23,Tabone23,Perotti23,Banzatti23,Gasman23b}. Our slab model fits do not provide tight constraints on the temperature of the emitting gas. This is likely due to the limits of fitting multiple transitions with a single slab model. As demonstrated by \citet{Banzatti23a} and \citet{Banzatti23}, \water\ lines close together in frequency space can originate from physically distinct locations, with distinct densities and temperatures. In the future more sophisticated modeling, e.g., using nested slab models or full radiative transfer of a model disks, will be needed to provide tighter constraints.

As previously mentioned, CO ro-vibrational emission at 4.7 \um\ was not detected in ground-based observations of SY Cha \citep{Brown13}. However, ro-vibrational CO emission is clearly seen in our MIRI-MRS Band 1 data, which begin at 4.9 \um. 
Given the presence of extended emission seen in \hh, it is possible that the lower J CO lines are obscured by an absorbing disk wind between the disk and the viewer. Previous high-spectral resolution ground-based studies have shown that lower J CO lines are more likely to be seen in absorption than higher J lines, particularly for disks such as SY Cha's with inclinations greater than 40\degree\ \citep{Banzatti22}. This is interpreted as due to a disk wind component. Alternatively the 4.7 \um\ emission could be hidden by telluric features in the ground-based observations, as suggested by \citet{Brown13}.
Our best-fit slab model for CO gives an excitation temperature of 1570 K and an equivalent emission radius (emitting area) of 0.13 au (0.053 au$^2$). This is consistent with the temperatures and emitting area derived from LTE slab model fits to Keck-NIRSPEC data for full T Tauri disks by \citet{Salyk11b}. However, they found transition disks tended to be best fit by a lower temperature model and smaller emitting area. That the CO emission in SY Cha is consistent with emission from full protoplanetary disks suggests that the inner disk in SY Cha remains gas-rich. 

\thcotwo\ is not detected. Based on the best-fit temperature and column density for \cotwo, the \thcotwo\ feature at 15.42 \um\ is expected to be extremely weak, peaking at less than 1 mJy. This feature also overlaps with a bright \water\ line, preventing constraints on the \thcotwo/\cotwo\ ratio.

\begin{figure}
    \centering
    \includegraphics[width=\columnwidth]{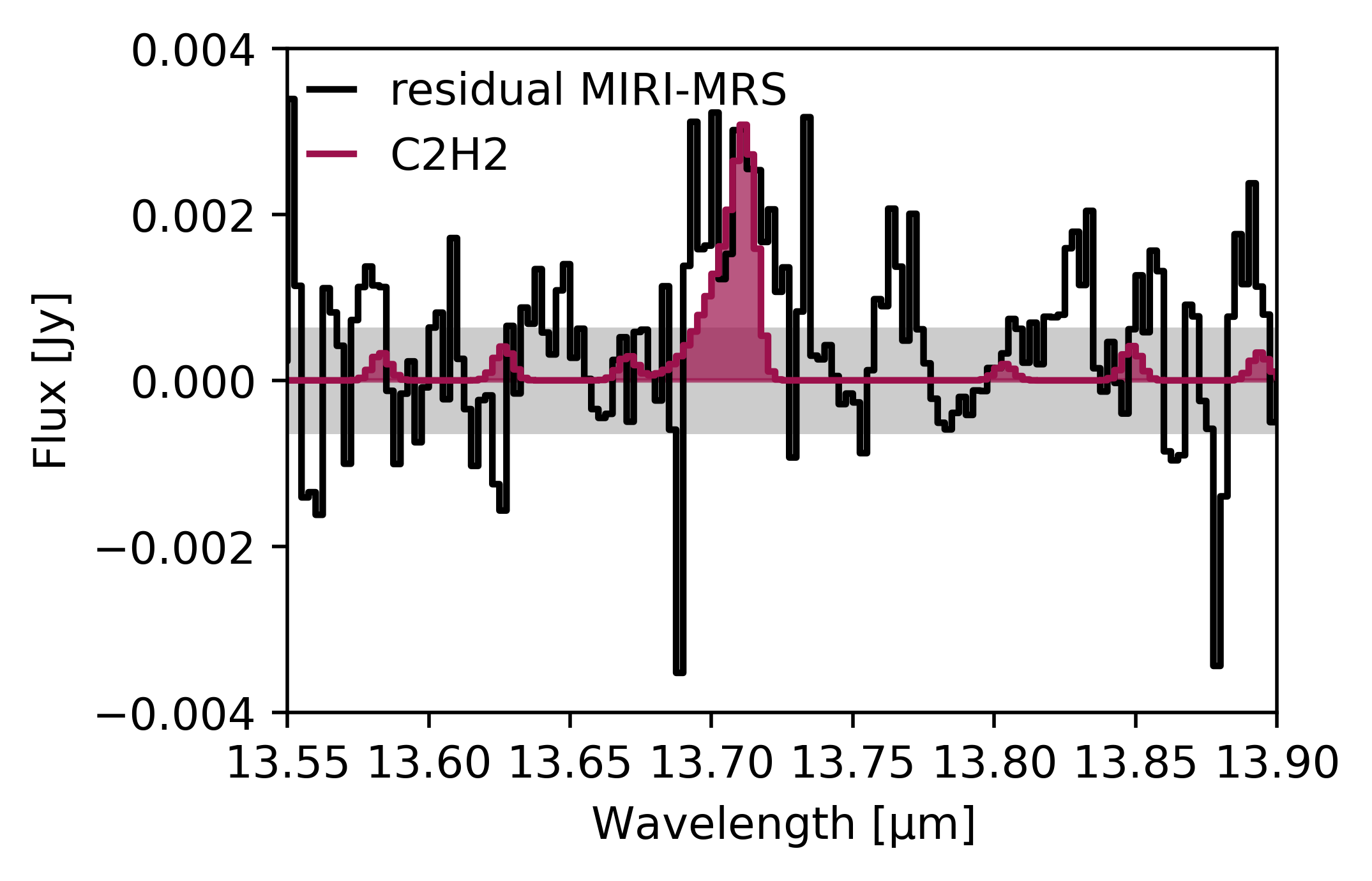}
    \caption{Residual observed emission after subtracting the continuum, \water, OH, \cotwo, and HCN best-fit models (black). The best-fit model for \cchh\ is overplotted in magenta. Grey shading indicated the $1\sigma$ rms noise level calculated for the continuum subtracted spectrum for 14.74-14.755 \um. The emission attributed to \cchh\ is of the same order as nearby residual emission.}
    \label{cchhres}
\end{figure}

Most mid-IR disk spectra show either strong \water\ emission, with \water\ emission stronger than emission from other molecular species, or hydrocarbon emission \citep{Pascucci13,Banzatti20,Grant23,Tabone23,Banzatti23,Gasman23b}. The SY Cha spectrum contains both strong \water\ lines as well as, potentially, a weak \cchh\ line. Figure~\ref{cchhres} shows the best-fit \cchh\ slab model as well as the residual observed spectrum after subtracting the best-fit \water, OH, \cotwo\ and HCN models. Emission from \cchh\ explains residual emission that is not well fit by the fundamental bending mode of HCN. This potential \cchh\ emission feature at 13.69 \um\ peaks $5\sigma$ above the rms noise of 0.6 mJy. However, this is of the same order as the residuals in other regions of the spectrum where we do not expect molecular emission. The column density and temperature of \cchh\ are also not well constrained by the slab model fits (Figure~\ref{chi2other}). We therefore classify the \cchh\ detection as tentative. Previously reported values for \cchh\ in SY Cha could have been contaminated by emission from other species due to the lower spectral resolution of \spitzer\ \citep{Salyk11,Banzatti20}. Alternatively, the \cchh\ column density in the inner disk has changed in the twelve years between observations. 

\begin{deluxetable}{cccc}
\tablewidth{\columnwidth}
\tablecolumns{2}
\tablecaption{Best-fit model parameters per spectral channel. The quoted uncertainties are the confidence intervals per fit parameter, given based on the $\chi^2$ maps and are included only for maps with a closed $1\sigma$ contour (see Figures~\ref{chi2water}-\ref{chi2other}).} For N the confidence intervals are given in log-space.
\label{slabprops}
\tablehead{
\colhead{Molecule} & \colhead{$N$} & \colhead{$T$} & \colhead{$R$} \\  
 & \colhead{[cm$^{-2}$]} & \colhead{[K]} & \colhead{[au]} 
}
\startdata 
\hline
\multicolumn{1}{c}{4.9-5.7 \um} & \multicolumn{3}{c}{$\sigma_{5.44-5.452} = 0.59$ mJy}\\
\hline
\water\ &   2.0\ee{19}  &   1020 &  0.026 \\
CO      &   3.0\ee{16}  &   1570 &  0.13  \\
\hline
\multicolumn{1}{c}{5.7-6.6 \um} & \multicolumn{3}{c}{$\sigma_{5.998-6.004} = 0.22$ mJy}\\
\hline
\water\ &   1.2\ee{20}$_{-0.7}^{+6.4}$   &   1080$_{-650}^{+580}$  &   0.022$_{-0.014}^{+0.42}$  \\
\hline
\multicolumn{1}{c}{6.6-7.6 \um}& \multicolumn{3}{c}{$\sigma_{6.89-6.90} = 0.40$ mJy}\\
\hline
\water\ &   2.0\ee{19}$_{-1.0}^{+4.7}$  &   970$_{-660}^{+850}$ &   0.029$_{-0.019}^{+0.51}$ \\
\hline
\multicolumn{1}{c}{13.5-15.5 \um}& \multicolumn{3}{c}{$\sigma_{14.74-14.755} = 0.62$ mJy}\\
\hline
\water\ &   8.4\ee{18}$_{-0.8}^{+1.0}$   &   660$_{-120}^{+130}$    &  0.17$_{-0.08}^{+0.35}$  \\
OH      &   3.3\ee{18}  &   1750$^b$  &  0.038   \\
\cotwo\ &   4.9\ee{17}  &   210  &  0.27   \\
HCN     &   2.7\ee{13}$_{-0.5}^{+4.4}$  &   720$_{-190}^{+265}$  &  6.9$_{-6.88}^{+3.07}$   \\
C$_2$H$_2^{a}$ & 2.3\ee{17} &   100  &  2.7 \\
\hline
\multicolumn{1}{c}{15.5-17.5 \um} & \multicolumn{3}{c}{$\sigma_{16.06-16.1} = 0.97$ mJy}\\
\hline
\water\ &   7.0\ee{18}$_{-0.7}^{+0.6}$ &   510$_{-65}^{+60}$ &   0.24$_{-0.08}^{+0.19}$ \\
OH  &   1.2\ee{18}  &   810$^b$ &   0.15 \\
\hline
\multicolumn{1}{c}{17.9-20.7 \um} & \multicolumn{3}{c}{$\sigma_{19.1-19.153} = 0.33$ mJy}\\
\hline
\water\ &   3.1\ee{21}  &   360 &   0.27 \\
OH  &   1.1\ee{17}   & 3500$^b$ &    0.060 \\
\hline
\enddata 
\tablecomments{
$^a$Tentative detection\\  
$^b$ Temperatures for OH reflect collisional or chemical pumping, not a physical gas temperature
}
\end{deluxetable}

\section{Discussion}
The strong \water\ flux relative to carbon bearing species suggests the inner disk of SY Cha is \water\ rich, with a sub-solar carbon to oxygen ratio ($C/O < 0.5$). 
The \water\ column density from our slab models ranges from 7\ee{18} to 3.1\ee{21} cm$^{-2}$ for the different wavelength ranges fitted. This is sufficient for \water\ to be self-shielded from dissociating UV emission \citep{Bethell09}. It is possible that the high \water\ column also shields other species from UV photons \citep[e.g.,][]{Bosman22b}, resulting in emission from both \water\ and carbon bearing species. The best fit temperatures for \water\ emission in our models also vary from several hundred Kelvin to over a thousand Kelvin, with the best fit temperature generally decreasing with increasing wavelength. This can be explained by the \water\ transitions at shorter wavelengths having preferentially higher excitation temperatures \citep{Banzatti23a}, and has been observed in several other protoplanetary disks \citep{Gasman23b,Banzatti23}. The emission at different excitation temperatures likely also probes different radial regions in the disk: a much smaller area of the disk is able to reach temperatures warmer than 1000~K compared to a few hundred Kelvin. Thus, the best-fit emitting area decreases as the best fit temperature increases.

Prompt emission from OH is the result of \water\ photodissociation by UV photons forming OH in a rotationaly excited state \citep{Harich00,vanHarrevelt00}. In this circumstance, OH emission should be seen throughout the wavelength range covered by MIRI-MRS, with OH lines shortward of 10 \um\ brighter than those at longer wavelengths \citep{Tabone21}. We do not detect OH emission at wavelengths shorter than 13 \um, nor do the detected OH quadruplets display the asymmetry expected from prompt emission. While weak OH emission at short wavelengths could be hidden in the forest of water lines, the lack of asymmetry in the observed quadruplets leads us to conclude that the observed OH emission is not primarily the result of UV dissociation of \water. Instead, OH is likely excited via collisions or chemical pumping. Similar results were found for the disk around Sz 98 by \citet{Gasman23b}.

The large cavity observed in the millimeter continuum has not resulted in a \water\ poor inner disk. This is not surprising given that the spatially resolved millimeter \co\ emission observed with ALMA does not show an inner cavity, suggesting that there are not large gaps in the radial gas distribution. SY Cha's infrared spectral index, as well as the presence of silicate emission, is also indicative of disk material at small radii \citep[e.g.,][]{Banzatti20}. 

Most mechanisms proposed to form substructures such as dust rings in protoplanetary disks do so by creating a pressure trap \citep{Bae23}. Pressure traps are able to stop the inward drift of large, millimeter sized, dust particles, which can lead to gaps seen in millimeter continuum emission. Pressure traps do not inhibit the movement of gas or the small, micron-sized, particles which are more coupled to the gas. Gas and small grains can instead be cleared by photo-evaporation or interaction with a perturber, such as a planet or an infall stream \citep{Lin79,Lubow94,Suriano19,Kuznetsova22}. 
Based on the high column density of molecules in the inner disk, particularly \water\, as well as the high infrared spectral index and resolved CO emission, we conclude that the mechanism which created the millimeter dust ring at 100 au has not prevented the inward motion of gas and sub-micron dust in SY Cha.
If the millimeter cavity in SY Cha has been created by a planet, such a planet would need to be massive enough to create a pressure trap but below the mass needed to clear a gap in the gas \citep{Kley12,Zhu12}.

\subsection{Comparison to PDS 70}
\begin{figure*}
    \centering
    \includegraphics[]{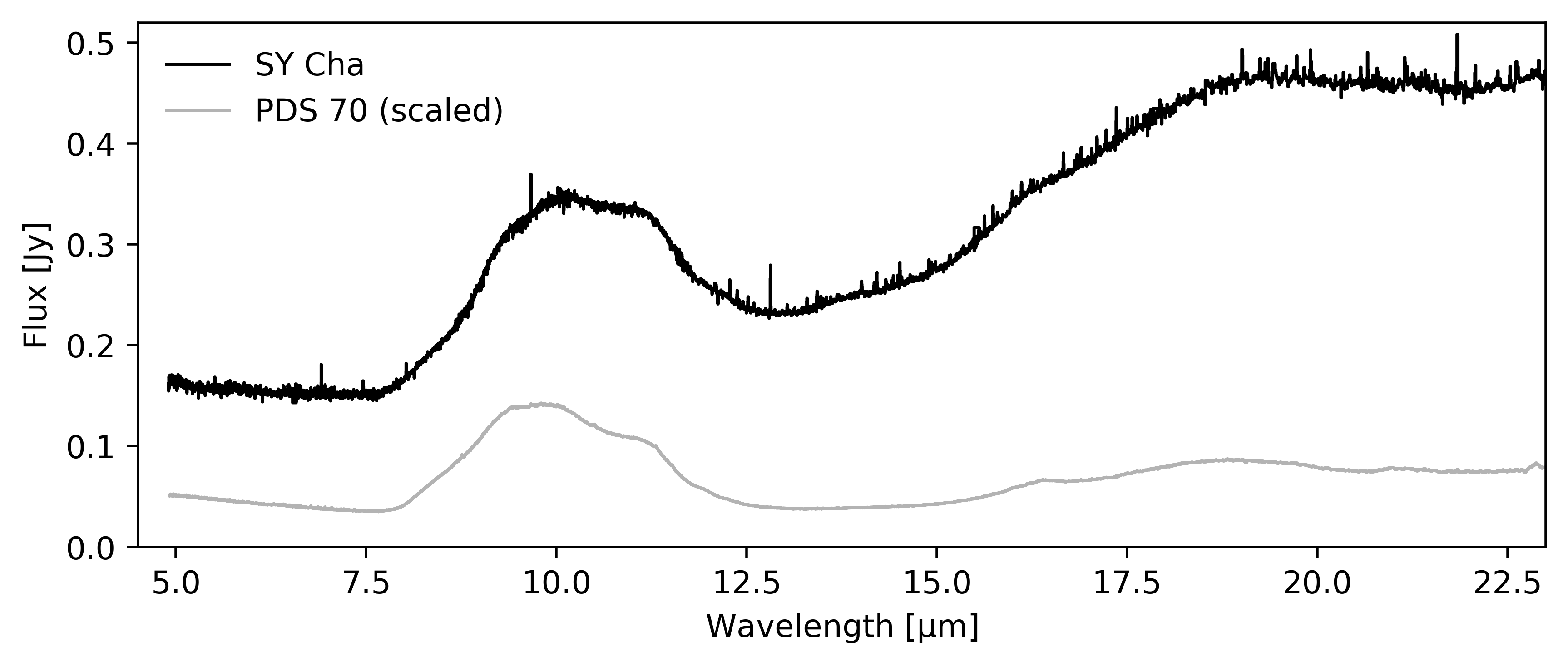}
    \caption{Comparison between the \jwst-MIRI MRS spectrum of SY Cha (black) and PDS 70 scaled to the distance of SY Cha (grey). SY Cha exhibits stronger flux in both the line and continuum emission.}
    \label{pds70}
\end{figure*}

The MIRI-MRS spectra of another transition disk, PDS 70, was recently studied by \citet{Perotti23}. The SY Cha and PDS 70 disks are superficially similar in the millimeter continuum, with a small inner disk, a wide ($>$40 au) dust cavity, an outer ring of millimeter grains, at roughly 70 au and 45 au respectively, and moderate inclinations of $51.1\degree$ and $51.7\degree$ \citep{Benisty21,Orihara23}.
Both disks have been observed with ALMA at $\sim20$ milli-arcsecond resolution, though the rms noise in the PDS 70 observations is significantly lower.
The inner disk of PDS 70 is marginally resolved by ALMA, extending out to 18 au, while the inner disk in SY Cha is unresolved but constrained to be smaller than 3.5 au. 
As previously mentioned, PDS 70 is one of the only transition disks known to host young massive planets within its dust cavity. 
However, the MIRI-MRS spectra of these two systems are vastly different, with the SY Cha continuum a factor of 3-6 stronger and much stronger molecular and atomic emission when PDS 70 is scaled to the distance of SY Cha (Figure~\ref{pds70}). The shape of the continuum is also different. The slope of the continuum at short wavelengths is extremely steep in PDS 70 when compared to most T Tauri disks.
The crystalline forsterite feature at 11 \um, indicative of a past heating event, is more pronounced in PDS 70, while the slope of the continuum beyond 15 \um\ is steeper for SY Cha.

The best-fit slab models for \water\ from 6.7-7.3 \um\ give similar temperatures and emitting areas for both sources, roughly 600 K and 0.04 au. However, the column density for SY Cha is several orders of magnitude greater, at 5.2\ee{21} \cmcm\ compared to 1.4\ee{18} \cmcm\ in PDS 70. Emission lines from other atomic and molecular species are likewise much stronger for SY Cha. 

There are several potential explanations for the different chemical composition in the inner disks of these two systems. With a bolometric luminosity of 0.35 \lsun, PDS 70 is less luminous than SY Cha \citep{Pecaut16}. Thus, the incident ionizing flux from the central star, a driver of disk evolution, is different. The inner disk in SY Cha could be less evolved, with less gas depletion, than the PDS 70 disk. Although the ages of these systems are highly uncertain, the best estimates suggest SY Cha is roughly 2 Myr younger than PDS 70 \citep{Muller18,Galli21}.

Alternatively, it is possible that the dust cavity in SY Cha was not carved by massive planets, instead being formed by other methods \citep[e.g.,][]{Flock15} or smaller planets, which are less efficient at preventing the inward motion of gas and small dust grains. 
In the case of planets, the strength of the pressure trap increases with increasing planet mass, as well as depending on the disk temperature and viscosity \citep[e.g.,][]{Fung14,Kanagawa15a,AsensioTorres21}. 
There is no published data on SY Cha from optical/near-IR surveys searching for planets in transition disks. Likewise, the published observations of SY Cha with ALMA are not sensitive enough to detect emission from protoplanets of similar brightness to the ones detected in PDS 70. 
The \co\ gas observed by ALMA does not display a central cavity, while the less abundant \coo\ and \cooo\ isotopologues exhibit a ring-like morphology. In PDS 70 all molecular species, apart from the optically thick \co, observed with ALMA appear depleted inside of $0\farcs{2}$ and modeling of the disk finds a gas surface density minimum at 20 au \citep{Keppler19,Facchini21,Portilla23}.
The cavity in PDS 70 is clearly far more depleted than that in SY Cha.
This is likely the cause of the different chemical compositions in the inner disk. 

If the cavity in SY Cha is carved by less massive planets this would point to the composition of the inner disk being strongly linked to the mass of, not just the presence of, planets at larger radii. 
Confirmation of this idea will require additional \jwst\ observations of transition disks to constrain the inner disk composition as well as deep searches for sub-stellar companions. 
Overall, the molecular column densities in the inner disk of SY Cha are more similar to those derived for the full protoplanetary disk of Sz 98 despite its different millimeter dust morphology \citep{Gasman23b}. 
The Sz 98 disk is a full protoplanetary disks with several narrow gaps observed in the sub-millimeter continuum. Its dust disk extends to $R=278$ au, compared to $\sim 140$ au for the SY Cha disk  \citep{Tazzari17,Orihara23}. 
 
It is also possible that the strong molecular emission lines in SY Cha are indicative of some level of dust grain growth in the inner disk. As dust grains grow, the dust opacity is reduced, exposing a larger gas column and resulting in stronger molecular line emission without a large change in the SED at infrared wavelengths \citep{Antonellini23}. In this case the unresolved inner disk emission seen with ALMA should be from dust emission, rather than free-free emission or an ionized wind. Additional observations at longer wavelengths would help clarify the origin of the sub-millimeter emission. It should be noted that the \jwst\ data presented in the current work demonstrates the presence of small dust grains in the inner disk but does not rule out the presence of larger grains. 

\subsection{Temporal Variability}
As discussed in Section~\ref{results:cont}, there is substantial variability observed in the continuum flux between the \jwst\ observations and the three \spitzer\ epochs.
To determine whether the molecular line flux is likewise variable we degrade the MIRI-MRS spectrum to the resolution of the \spitzer\ high resolution observations from 2004 and 2008. Continuum subtraction is then carried out as described in Section~\ref{reduction}, with the same points used for all epochs. We then integrate over the same wavelength range to get the total line flux. The wavelength ranges used for \cchh, HCN, and \cotwo\ are the same as those used by \citet{Banzatti20} while the wavelength range for \water\ is taken from \citet{Najita13}. The results are given in Table~\ref{linecompare}.

\begin{figure*}
\gridline{\fig{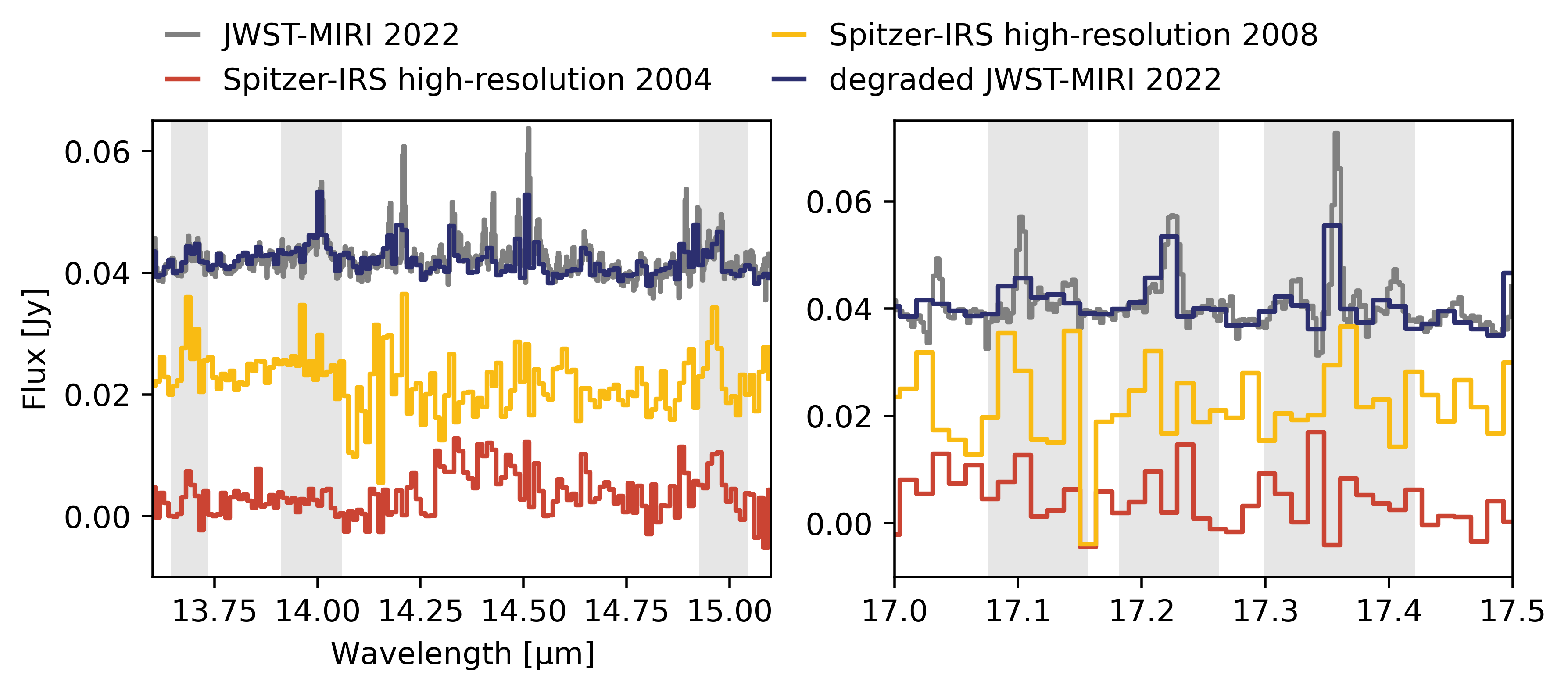}{0.9\textwidth}{}}
\caption{Continuum subtracted spectra from \spitzer\ high resolution and MIRI MRS observations, with the MIRI spectra degraded to the same resolution as \spitzer\ before continuum subtraction. Epochs are offset by 0.02 Jy for readability. The grey shaded regions indicate the wavelength range used when integrating line fluxes. The line flux shows noticeable variation between epochs, though the changes are within the flux calibration uncertainty.}
\label{epochs}
\end{figure*}

The total line flux appears to vary not just between the \jwst\ and \spitzer\ observations, but also between the two \spitzer\ epochs. This apparent change in brightness is of the same order as that reported  by \citet{Banzatti12} for EX Lupi for similar wavelength ranges. EX Lupi is the prototype of the highly variable T Tauri sources know as EXors and the changes were seen when comparing observations during quiescent and outbursting epochs. However, the luminosity of the line emission in SY Cha is weaker than that in EX Lupi. 

To determine if the variability seen in the molecular line emission for SY Cha is significant, we calculate the uncertainty in the flux ratio between two epochs, where the uncertainty for a single observation is the calculated rms noise added in quadrature to the spectro-photometric accuracy of the instrument. 
Given the spectro-photometric accuracy of both \spitzer-IRS and \jwst-MIRI, the variability in line flux in SY Cha is not statistically significant. The potential variability in line emission seen in SY Cha, which has not undergone any known recent outbursts, suggests that the physical conditions in the inner regions of even quiescent protoplanetary disks could change on observable timescales. Careful study of line emission in a single source at high signal-to-noise over multiple epochs is needed to confirm this idea.

\begin{deluxetable*}{cccccccc}
\tablewidth{0pt}
\tablecolumns{1}
\tablecaption{Comparison of Integrated Line Flux in \spitzer\ and \jwst\ Observations. Error is the rms noise error.}
\label{linecompare}
\tablehead{
\colhead{Molecule} & \colhead{Wavelength} & \colhead{\spitzer\ 2004} & \colhead{Err} & \colhead{\spitzer\ 2008} & \colhead{Err}& \colhead{\jwst\ 2022}  & \colhead{Err}\\  
\colhead{ } & \colhead{\um} & \twocolhead{(\eten{-14} erg s$^{-1}$ cm${-2}$)} & 
\twocolhead{(\eten{-14} erg s$^{-1}$ cm${-2}$)} & 
\twocolhead{(\eten{-14} erg s$^{-1}$ cm${-2}$)}
}
\startdata 
\cchh\ & 13.648-13.729  & $0.28$ & $0.18$ & $0.66$ & $0.25$ & $0.22$ & $0.14$\\ 
HCN     & 13.914-14.055 & $0.29$ & $0.34$ & $1.12$ & $0.47$ & $0.96$ & $0.27$\\ 
\cotwo\ & 14.93-15.04  & $0.51$ & $0.21$ & $0.51$ & $0.29$ & $0.28$ & $0.17$\\  
\water\ & 17.077-17.1555 & $0.45$ & $0.15$ & $0.18$ & $0.25$ & $0.156$ & $0.094$ \\
\water\ & 17.183-17.261 & $0.32$ & $0.11$ & $0.22$ & $0.19$ & $0.258$ & $0.073$ \\
\water\ & 17.2998-17.4200 & $0.59$ & $0.23$ & $0.37$ & $0.38$ & $0.13$ & $0.15$ \\
\enddata 
\end{deluxetable*}

\section{Conclusions}
We report the first results from \jwst-MIRI MRS observations of the SY Cha protoplanetary disk, observed as part of the MINDS GTO program. We detect emission from HI, [NeII], \hh, \water, OH, CO, \cotwo, and HCN as well as a marginal detection of \cchh. 
The line fluxes for all species are relatively high compared to the other T Tauri disks observed thus far with MIRI-MRS, with the emission from \water\ particularly strong \citep{Grant23,Perotti23,Banzatti23,Gasman23b}. Based on these results we conclude that the high molecular column densities in the inner disk are likely due to the inward drift of gas and small grains while large grains are caught by a weak pressure trap. 
The superior spectral resolution of MIRI-MRS allows us to isolate the mid-infrared OH lines in the source for the first time. Based on the lack of OH emission observed shortward of 13 \um\ and the lack of asymmetry in the observed quadruplets, we conclude OH is not excited when \water\ is dissociated by UV photons. 
We also find that SY Cha is highly variable in the mid-IR continuum. This is consistent with previous studies of protoplanetary disks, particularly those with large millimeter dust cavities.
We further find hints of molecular line variability. Higher signal-to-noise observations will be needed to confirm this.

Determining how common gas-rich inner disks co-existing with large millimeter cavities are will require a large \jwst\ survey of protoplanetary disks with a range of dust morphologies. Further, 
multiple epochs of high spectral resolution observations with MIRI-MRS will be required to determine if variability of both line and continuum flux is common, while modeling will be needed to determine if this is due to changes in chemical abundance, temperature, or disk morphology. 

\section{Acknowledgments}
This work is based on observations made with the NASA/ESA/CSA James Webb Space Telescope. The data were obtained from the Mikulski Archive for Space Telescopes at the Space Telescope Science Institute, which is operated by the Association of Universities for Research in Astronomy, Inc., under NASA contract NAS 5-03127 for JWST. 
These observations are associated with program 1282. The following National and International Funding Agencies funded and supported the MIRI development: 
NASA; ESA; Belgian Science Policy Office (BELSPO); Centre Nationale d’Etudes Spatiales (CNES); Danish National Space Centre; Deutsches Zentrum fur Luft und Raumfahrt (DLR); Enterprise Ireland; 
Ministerio De Econom\'ia y Competividad; 
Netherlands Research School for Astronomy (NOVA); 
Netherlands Organisation for Scientific Research (NWO); 
Science and Technology Facilities Council; Swiss Space Office; 
Swedish National Space Agency; and UK Space Agency.

The data presented in this article were obtained from the Mikulski Archive for Space Telescopes (MAST) at the Space Telescope Science Institute. The specific observations analyzed can be accessed via \dataset[DOI]{https://doi.org/10.17909/8ref-ye65}.

K.S. and T.H. acknowledge support from the European Research Council under the Horizon 2020 Framework Program via the ERC Advanced Grant Origins 83 24 28. 
G.P. gratefully acknowledges support from the Max Planck Society.
D.B. and M.M.C. have been funded by Spanish MCIN/AEI/10.13039/501100011033 grants PID2019-107061GB-C61 and No. MDM-2017-0737. 
I.K., A.M.A., and E.v.D. acknowledge support from grant TOP-1 614.001.751 from the Dutch Research Council (NWO). 
I.K. and J.K. acknowledge funding from H2020-MSCA-ITN-2019, grant no. 860470 (CHAMELEON).
E.v.D. acknowledges support from the ERC grant 101019751 MOLDISK and the Danish National Research Foundation through the Center of Excellence ``InterCat'' (DNRF150). 
A.C.G. acknowledges from PRIN-MUR 2022 20228JPA3A “The path to star and planet formation in the JWST era (PATH)” and by INAF-GoG 2022 “NIR-dark Accretion Outbursts in Massive Young stellar objects (NAOMY)” and Large Grant INAF 2022 “YSOs Outflows, Disks and Accretion: towards a global framework for the evolution of planet forming systems (YODA)”.
V.C. acknowledges funding from the Belgian F.R.S.-FNRS.
D.G. thanks the Belgian Federal Science Policy Office (BELSPO) for the provision of financial support in the framework of the PRODEX Programme of the European Space Agency (ESA).
T.P.R. acknowledges support from ERC grant 743029 EASY.
D.R.L. acknowledges support from Science Foundation Ireland (grant number 21/PATH-S/9339).
M.T. acknowledges support from the ERC grant 101019751 MOLDISK.

\facility{JWST, Spitzer}
\software{\textsc{emcee} \citep{emcee}, \textsc{matplotlib} \citep{matplotlib}, \textsc{numpy} \citep{numpy}, \textsc{spectres} \citep{Carnall17}, \textsc{VIP} \citep{VIP1,VIP2}}

\begin{appendix}
\section{Continuum Subtraction}\label{app:contsub}
In Table~\ref{conttab} we list the points used to define the spline fit used to subtract continuum.

\begin{deluxetable}{ll}[!h]
\tablewidth{\columnwidth}
\tablecolumns{2}
\tablecaption{Points used in continuum subtraction.}
\label{conttab}
\tablehead{
\colhead{Covered Transitions} & \colhead{Value}  
}
\startdata
CO, \water\ & 4.926, 4.947, 4.987, 5.1, 5.168, 5.338, 5.439, 5.575, 5.702, 5.73 \\
HI 10-6 & 5.073, 5.1, 5.168, 5.1925, 5.225 \\
\water\ & 5.728,5.893,6.0,6.084,6.164,6.18,6.192,6.21,6.429,6.463,6.607 \\
\water\ & 6.608,6.761,6.895,7.159,7.267,7.444,7.56,7.626 \\
HI 12-7 & 6.749, 6.782, 6.7896, 6.801 \\
HI 6-5  &   7.4, 7.4250, 7.4950, 7.528 \\
HI 8-6, HI 11-7   &   7.425, 7.4775, 7.492, 7.528 \\
$\mathrm{[NeII]}$ 2P$_{1/2}$-2P$_{3/2}^a$ & 12.77,12.802,12.860,12.88\\
\water, OH, CO$_2$ \cchh, HCN & 
13.52,13.59,13.746,14.101,14.573,14.7,15.208,15.393,15.5 \\
\water, OH & 15.565,15.71,15.826,15.905,16.084,16.36,16.675,17.075,17.367,17.43,17.48,17.592,17.823\\
\water, OH & 17.876,18.299,18.615,18.735,19.125,19.40,19.9,20.21,20.36,20.685 \\
\enddata

\end{deluxetable}

\section{$\chi^2$ maps of fits}
\begin{figure}
\gridline{\fig{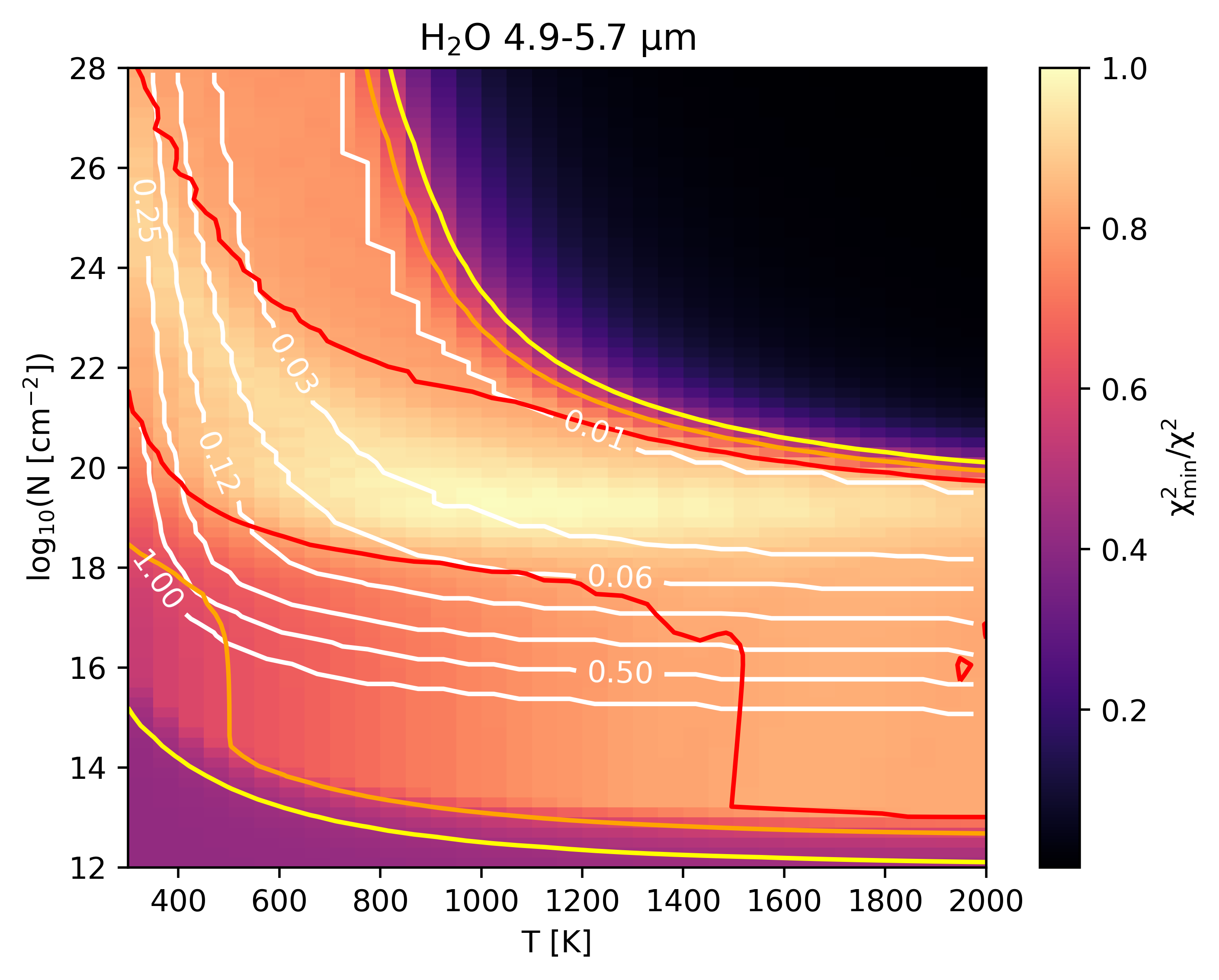}{0.45\textwidth}{(a)}
          \fig{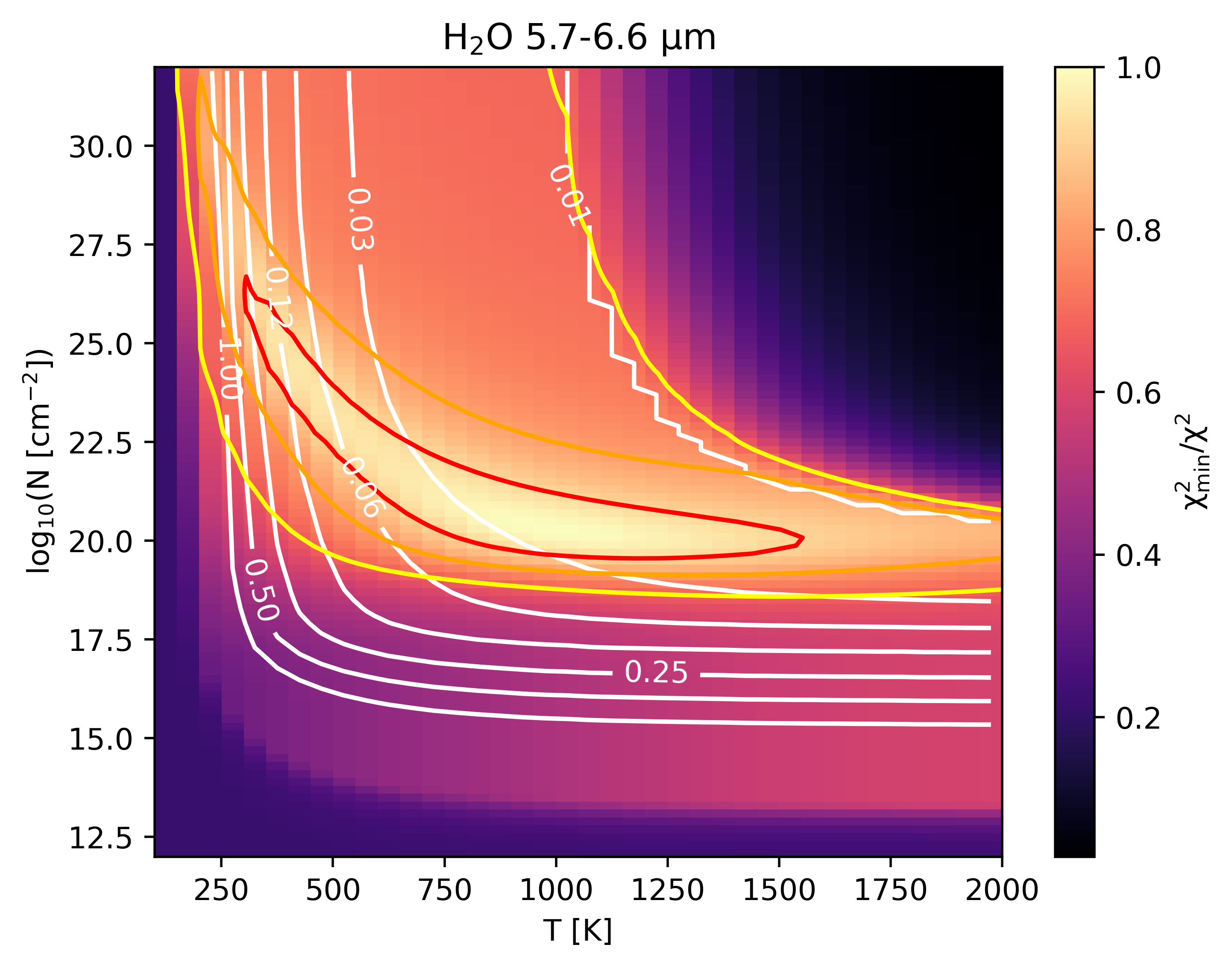}{0.45\textwidth}{(b)}}
\gridline{\fig{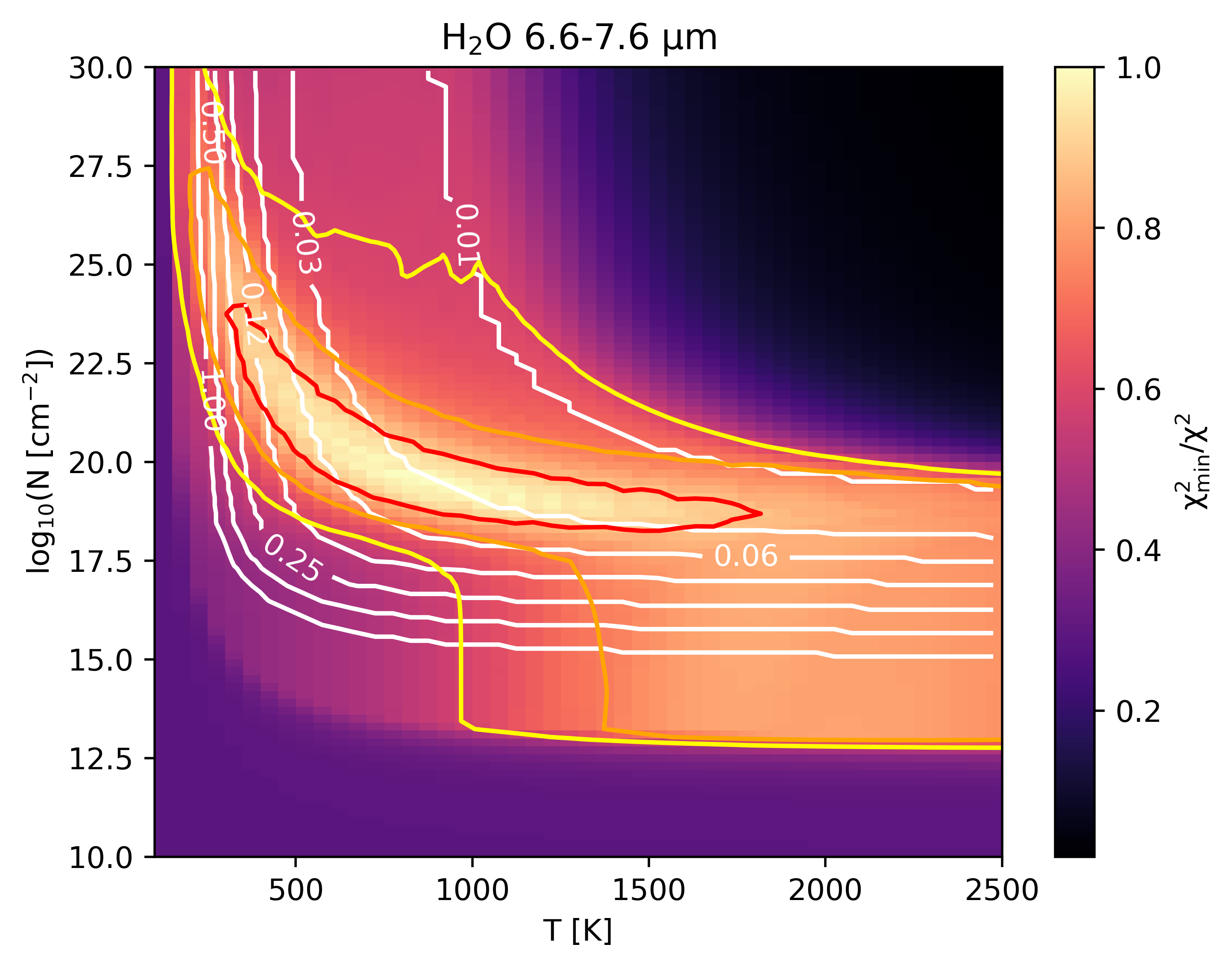}{0.45\textwidth}{(c)}
          \fig{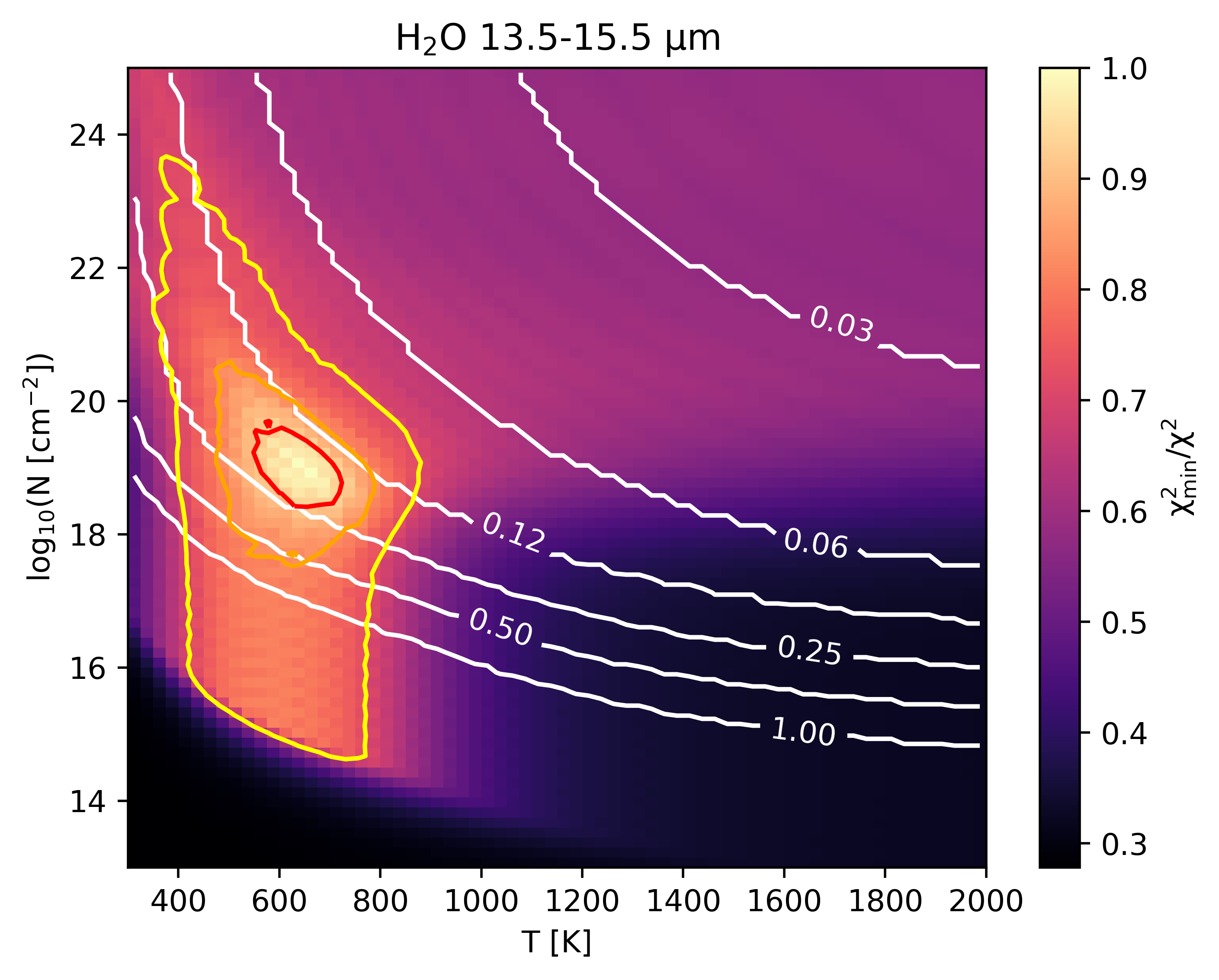}{0.45\textwidth}{(d)}}
\gridline{
          \fig{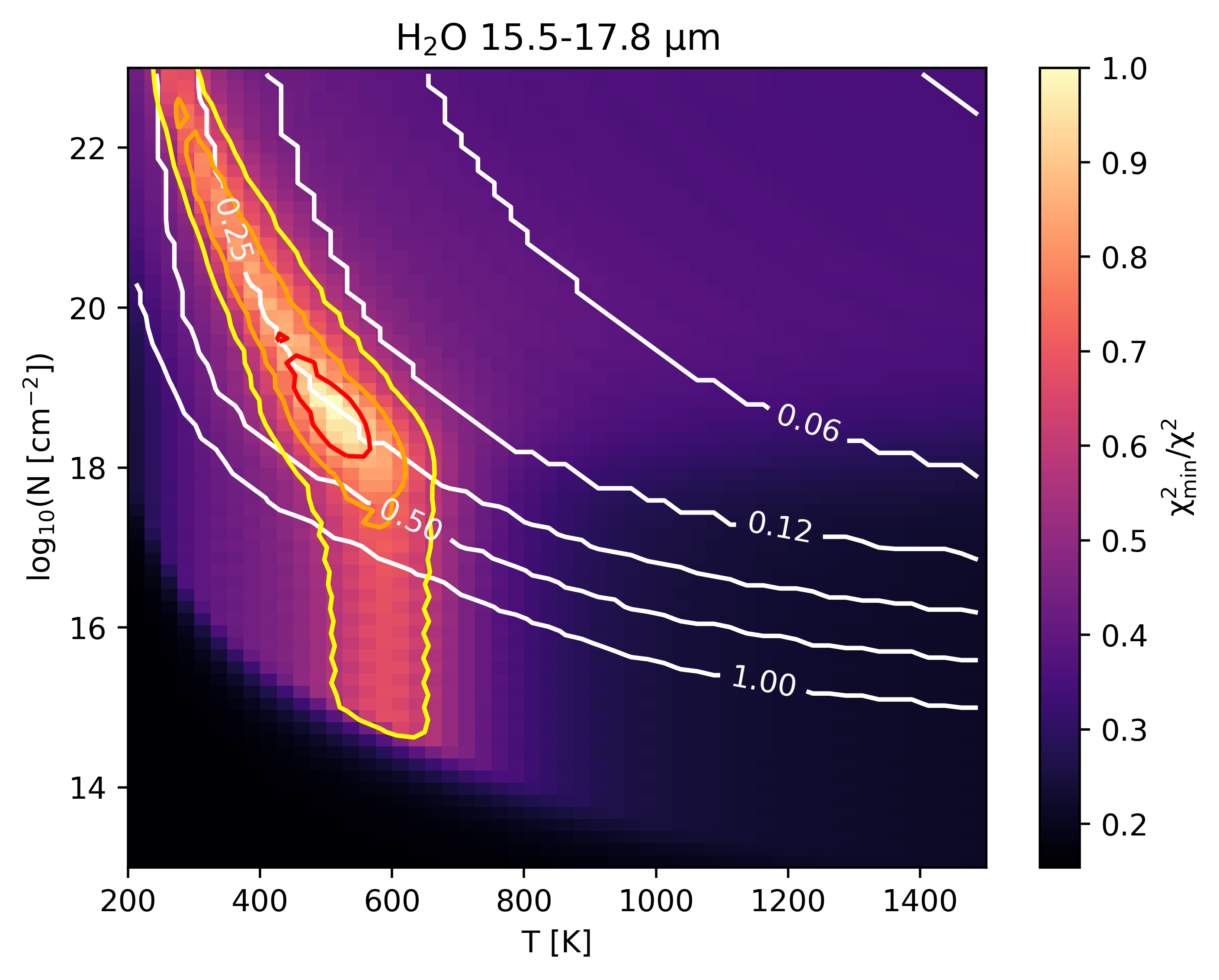}{0.45\textwidth}{(e)}
          \fig{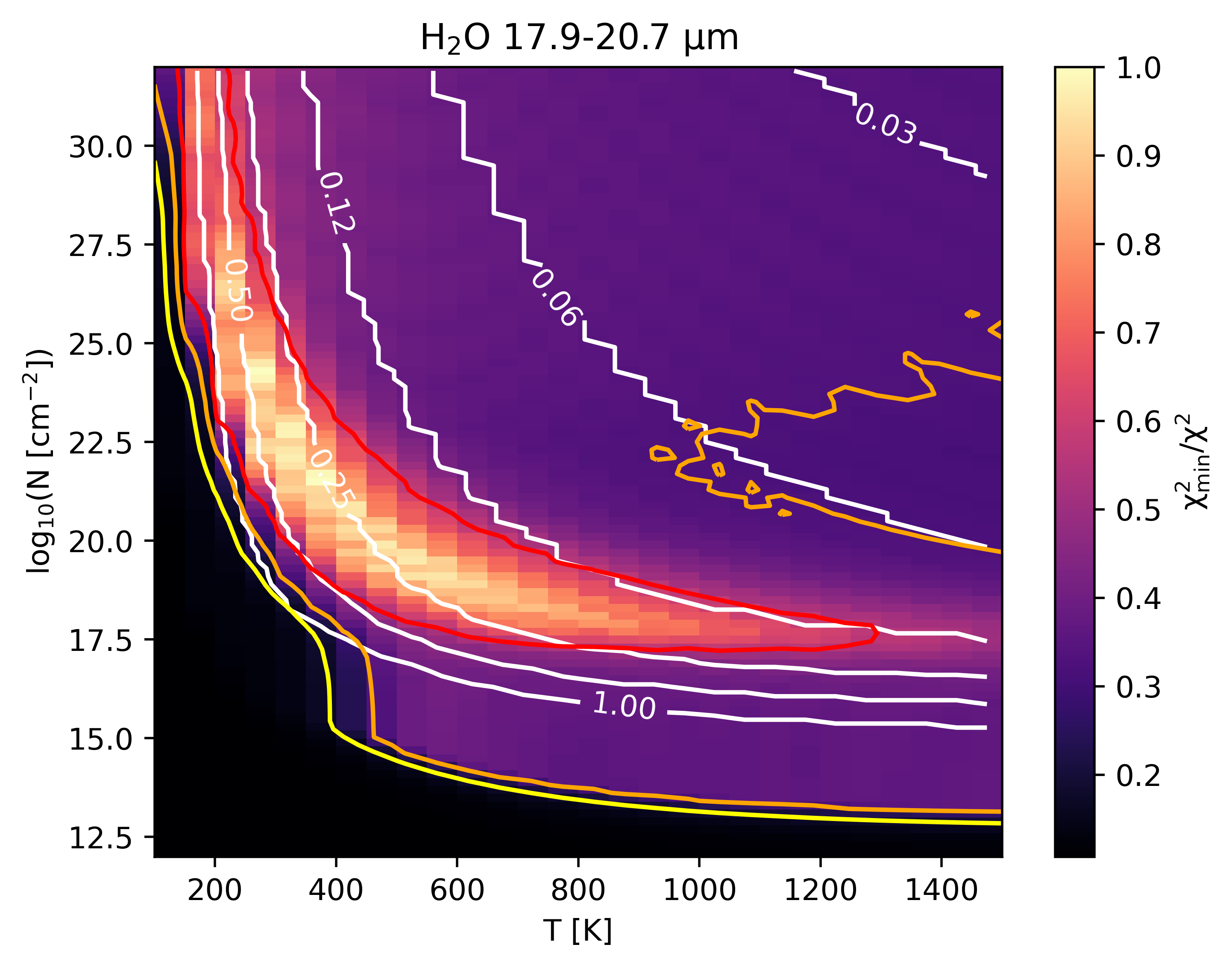}{0.45\textwidth}{(f)}}
\caption{$\chi^{2}$ maps for the slab model fits to \water. The color map shows $\chi^{2}_{min}/\chi^2$ such that the best-fit is given a value of 1. The red, orange, and yellow contours give the $1\sigma$, $2\sigma$, and $3\sigma$ uncertainty levels for 2 degrees of freedom. The white contours show the best-fit emitting radius, with labels in au.}
\label{chi2water}
\end{figure}

\begin{figure}
\gridline{\fig{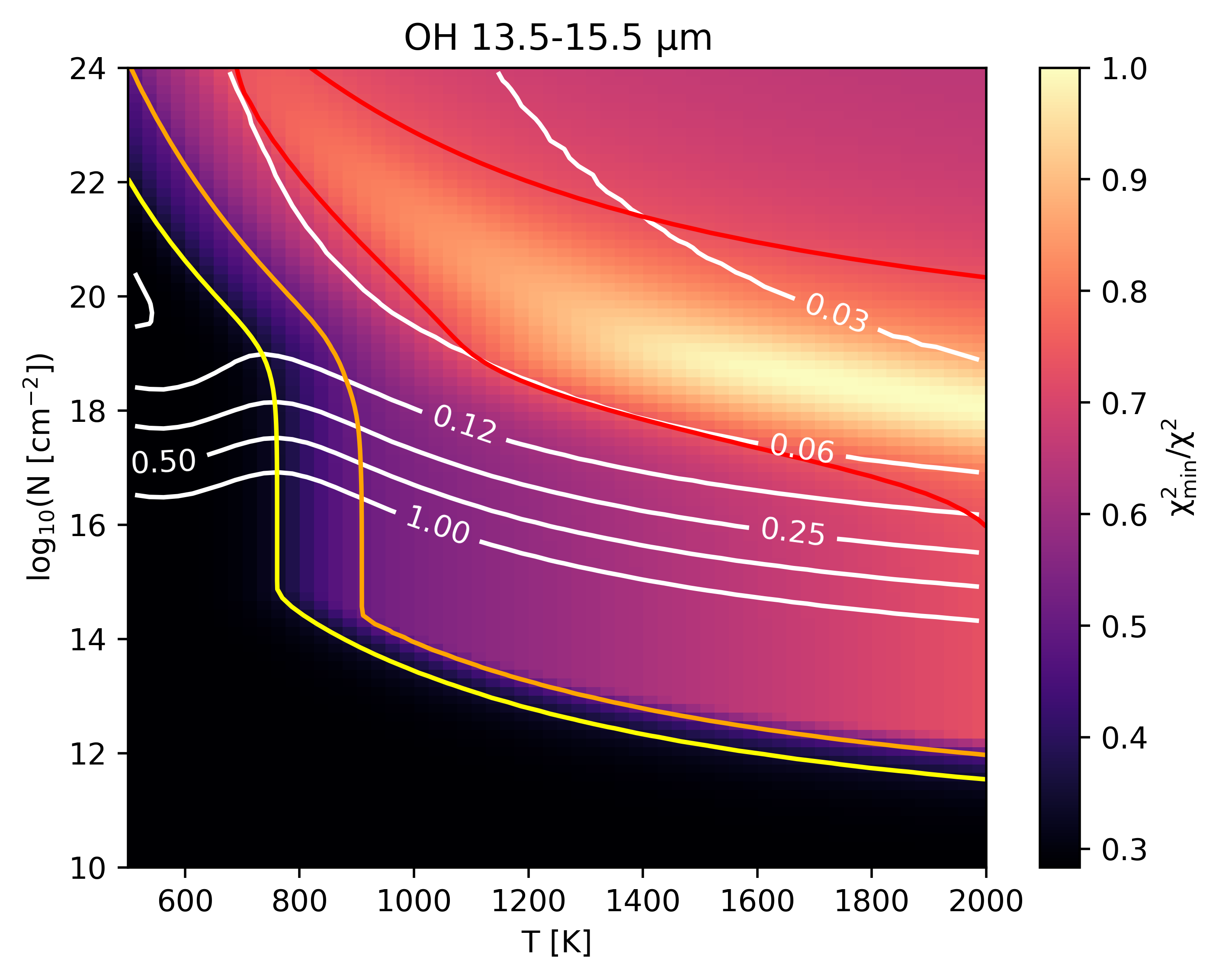}{0.45\textwidth}{(a)}
        \fig{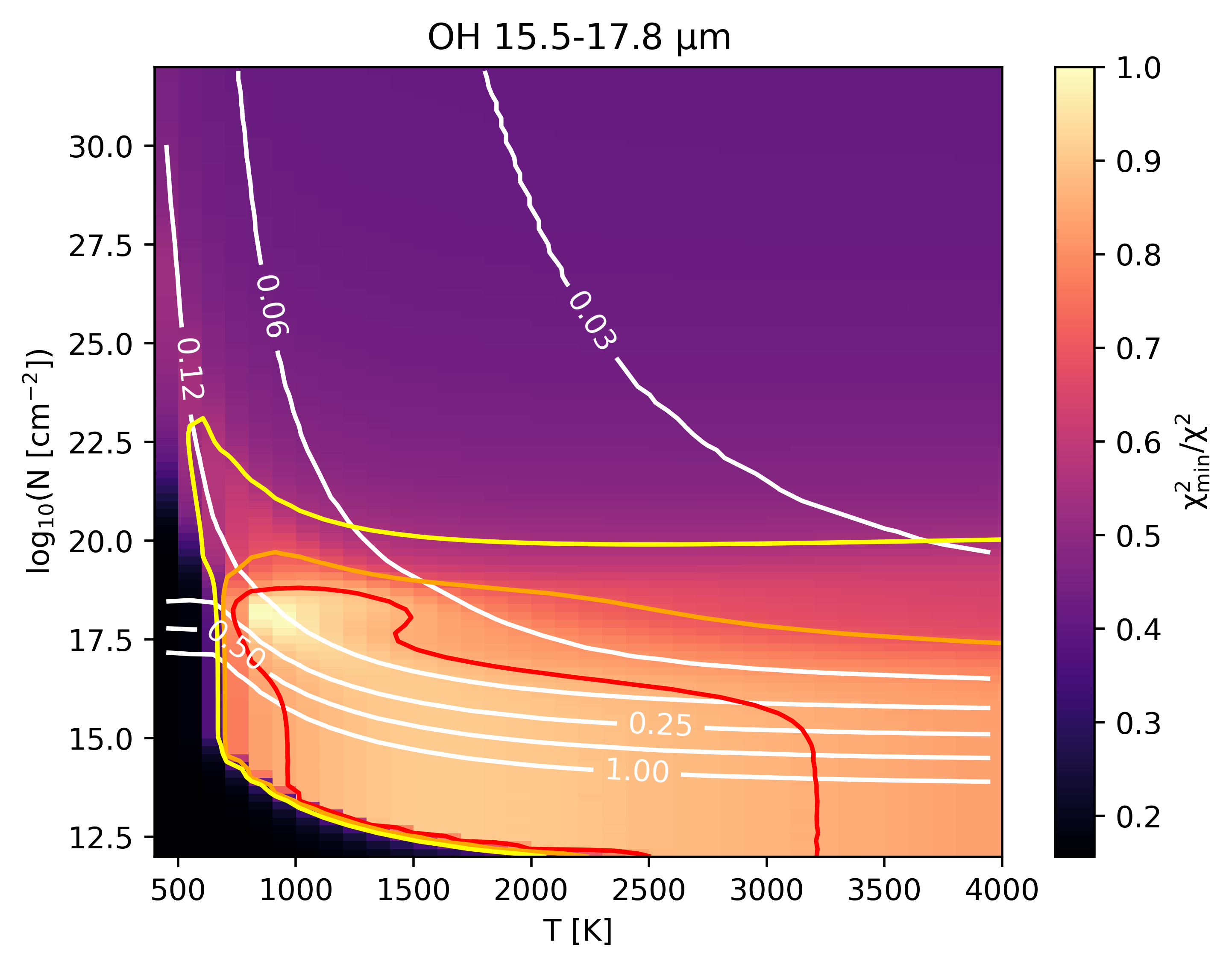}{0.45\textwidth}{(b)}}
\gridline{\fig{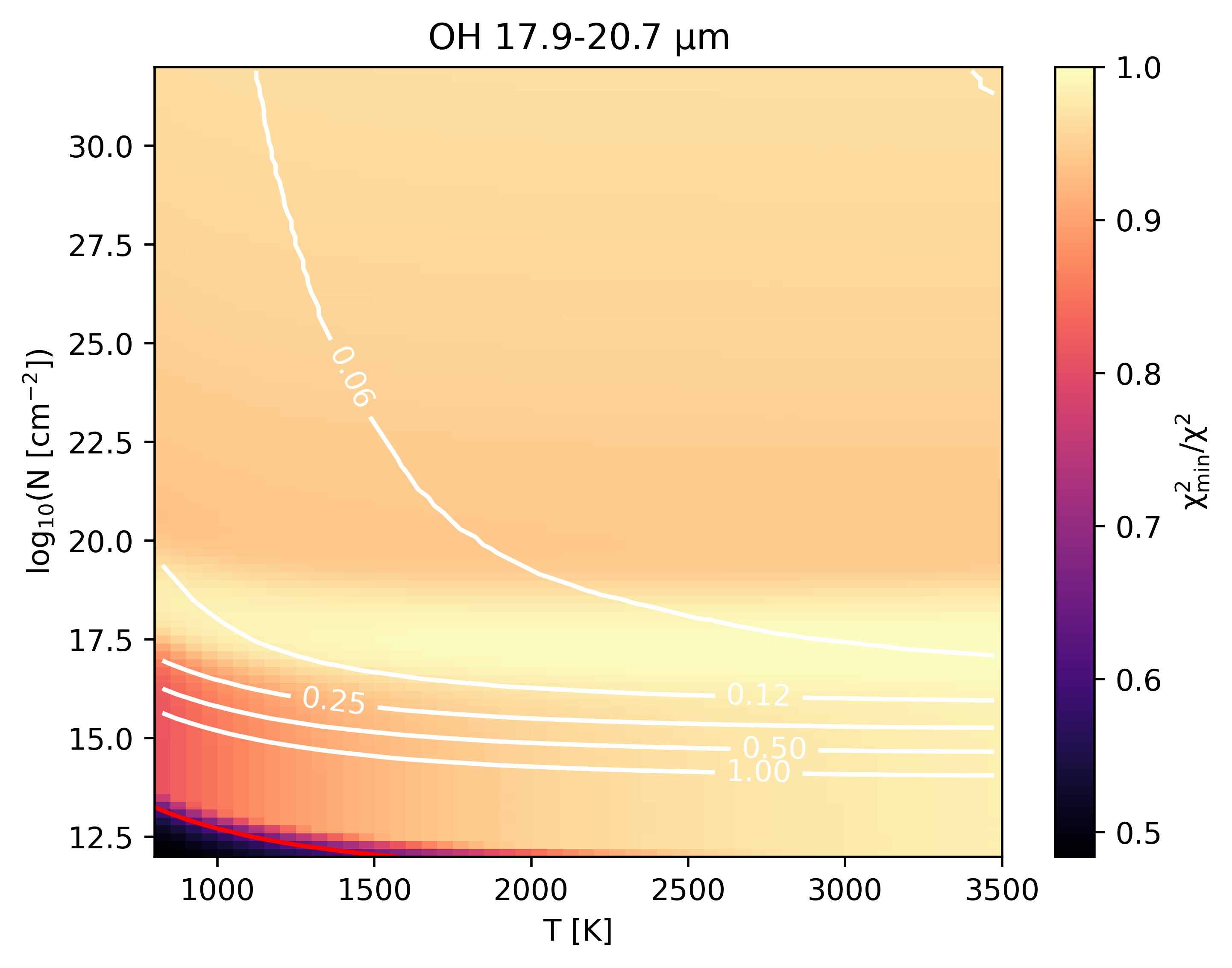}{0.45\textwidth}{(c)}}
\caption{$\chi^{2}$ maps for the slab model fits to OH. The color map shows $\chi^{2}_{min}/\chi^2$ such that the best-fit is given a value of 1. The red, orange, and yellow contours give the $1\sigma$, $2\sigma$, and $3\sigma$ uncertainty levels for 2 degrees of freedom. The white contours show the best-fit emitting radius, with labels in au. The OH lines are poorly constrained by slab models.}
\label{chi2oh}
\end{figure}

\begin{figure}
\gridline{\fig{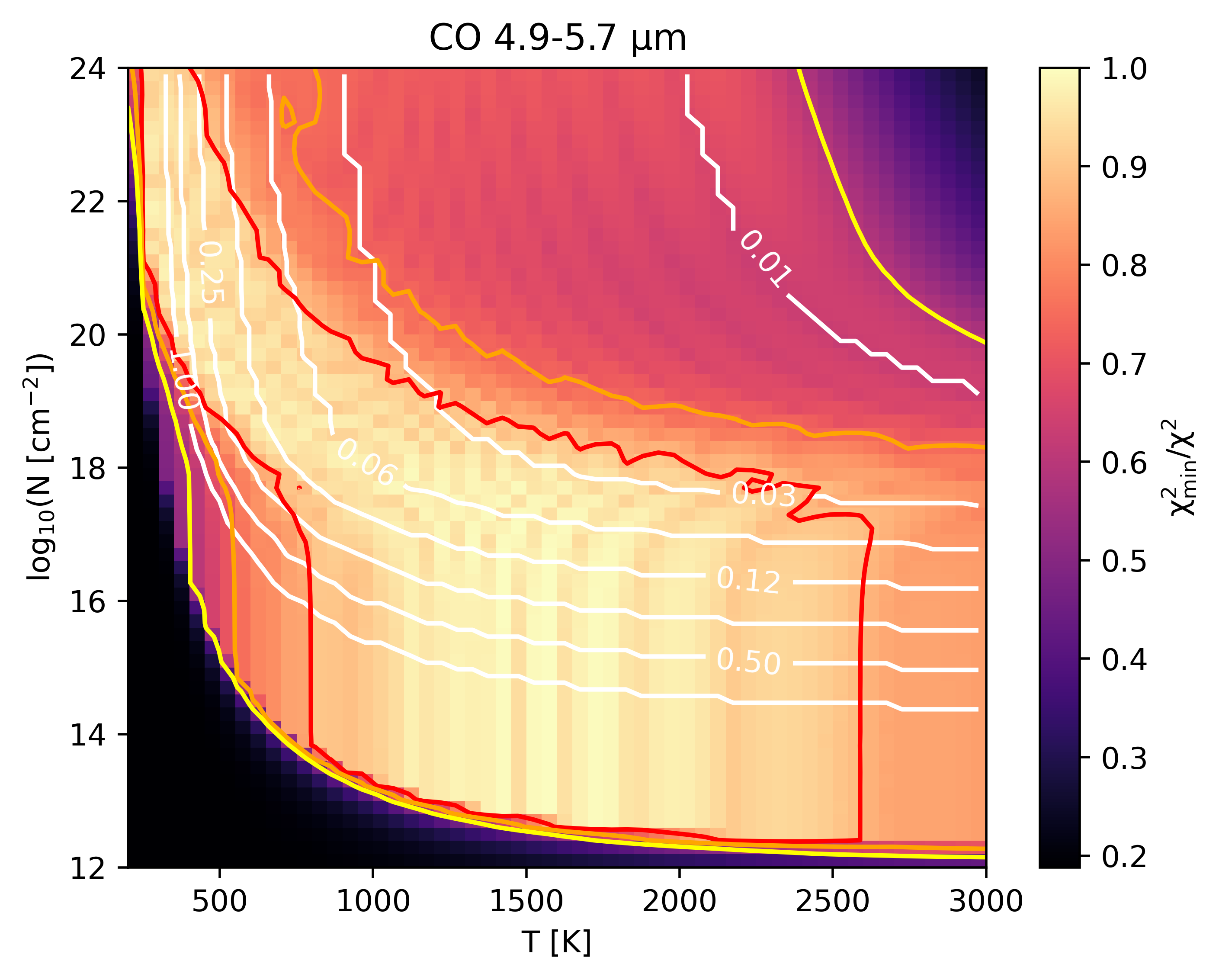}{0.45\textwidth}{(a)}
          \fig{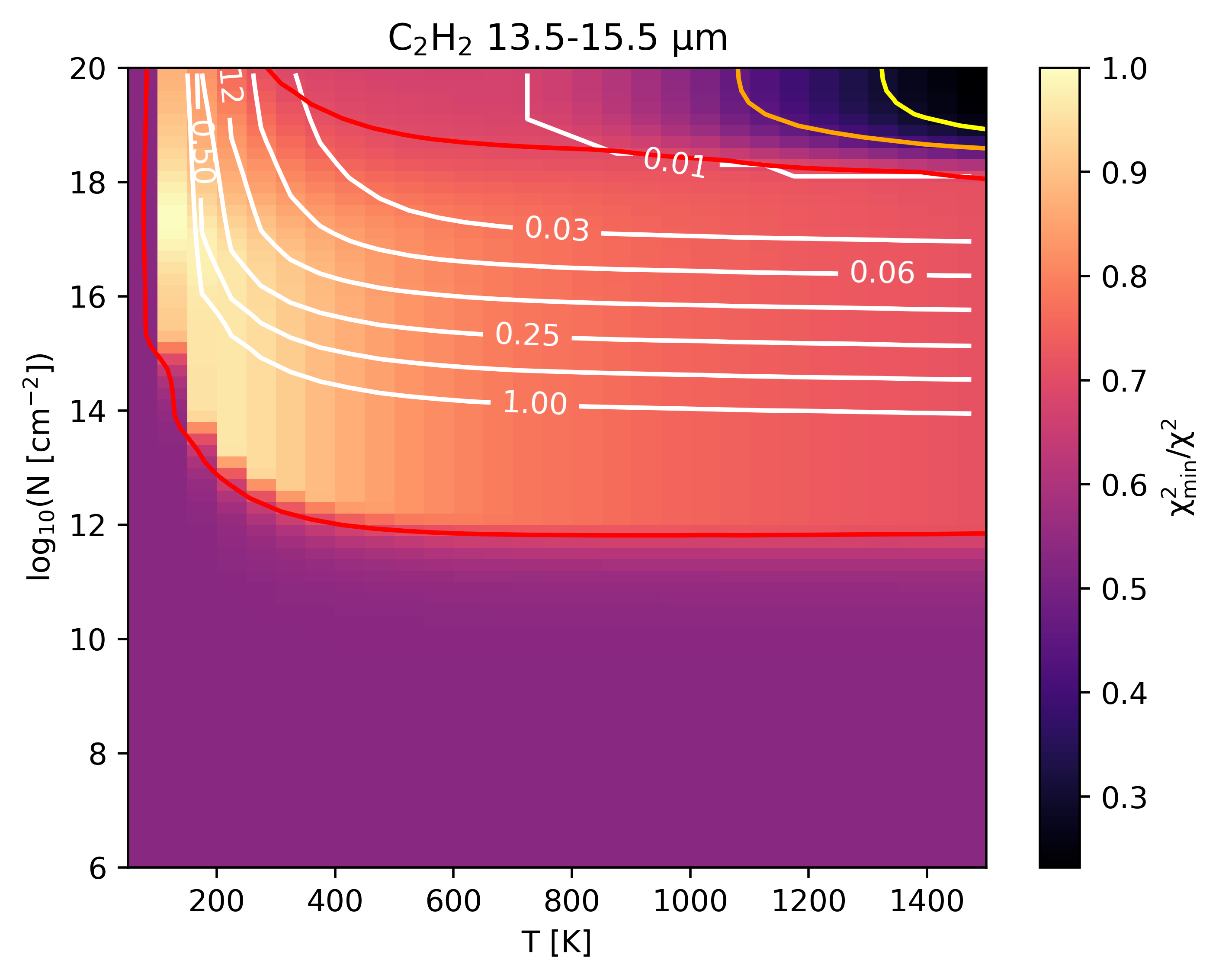}{0.45\textwidth}{(b)}}
\gridline{\fig{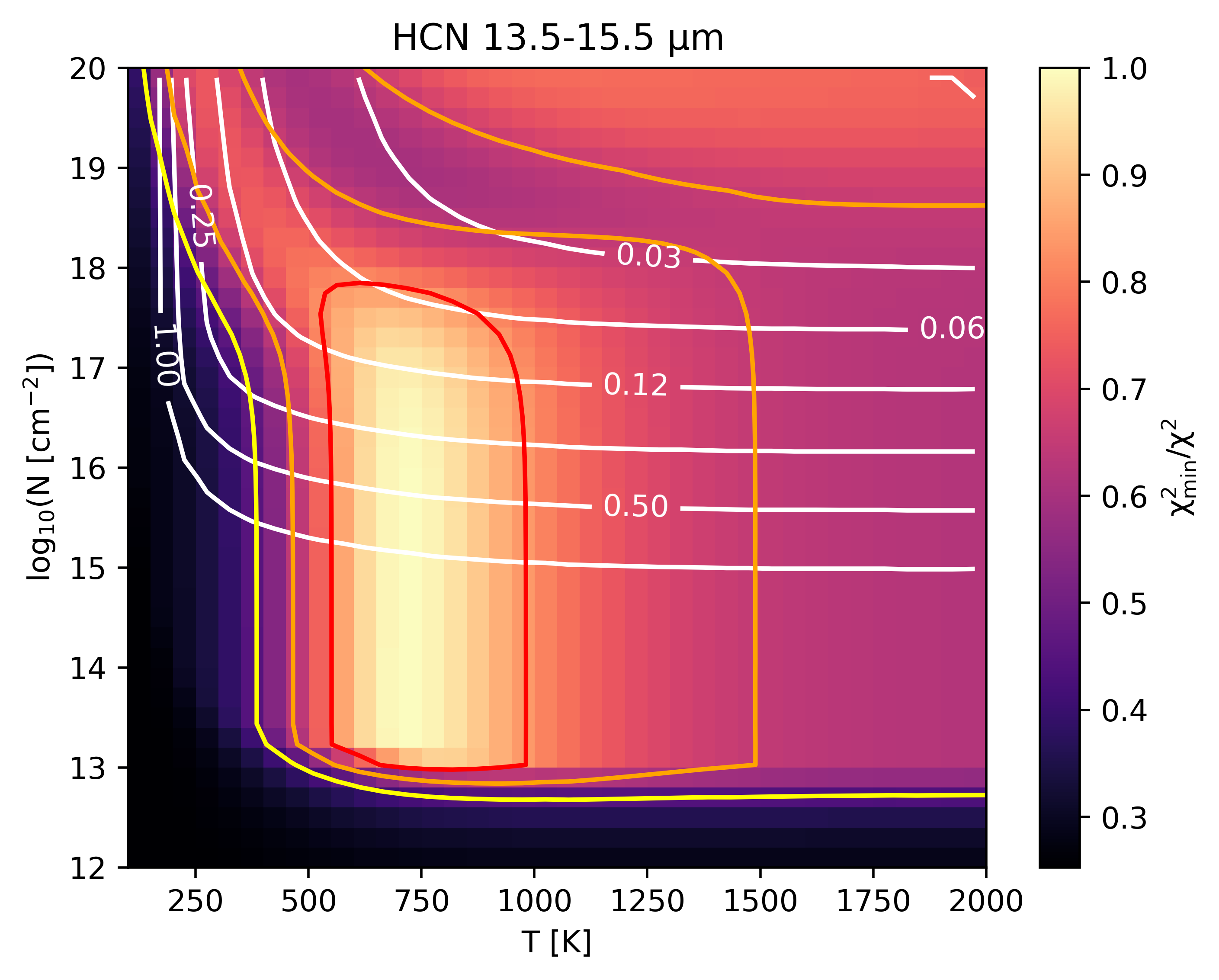}{0.45\textwidth}{(c)}
          \fig{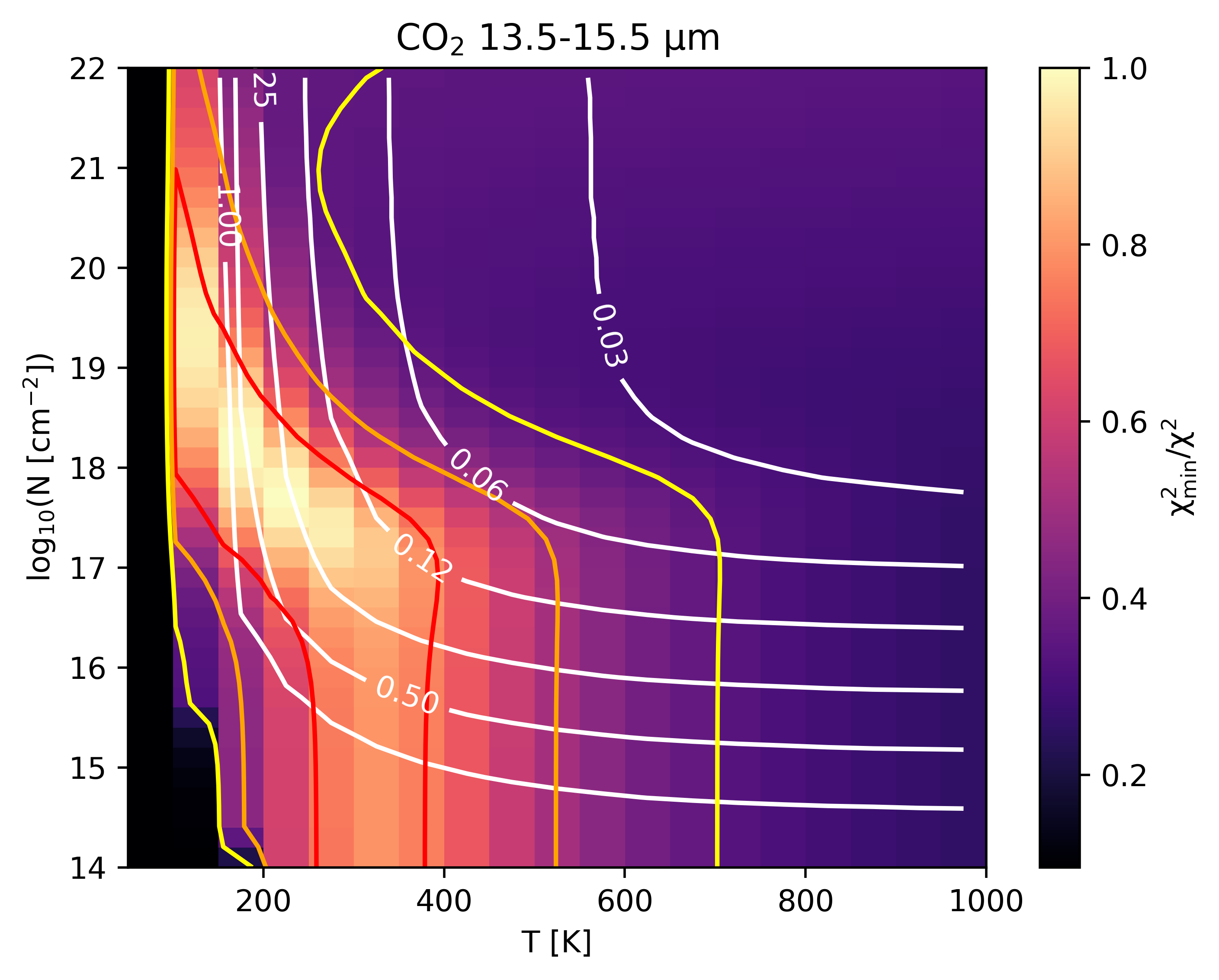}{0.45\textwidth}{(d)}}
\caption{$\chi^{2}$ maps for the slab model fits to CO, C$_2$H$_2$, HCN, and \cotwo. The color map shows $\chi^{2}_{min}/\chi^2$ such that the best-fit is given a value of 1. The red, orange, and yellow contours give the $1\sigma$, $2\sigma$, and $3\sigma$ uncertainty levels for 2 degrees of freedom. The white contours show the best-fit emitting radius, with labels in au.}
\label{chi2other}
\end{figure}

The $\chi^2$ per molecule and per fitted region are presented here (Figures~\ref{chi2water}-\ref{chi2other}).

\end{appendix}

\bibliography{main}{}
\bibliographystyle{aasjournal}

\end{document}